\documentclass[twocolumn,twocolappendix]{aastex6}

\usepackage{newtxtext,newtxmath}

\usepackage[T1]{fontenc}
\usepackage{ae,aecompl}

\usepackage{graphicx}	
\usepackage{epstopdf}
\usepackage{amsmath}	
\usepackage{amssymb}	
\usepackage{hyperref}
\usepackage{longtable}
\usepackage{comment}

\newcommand{\kd}{KMOS$^{\rm 3D}$\,}

\begin{document}

\defcitealias{WuytsS16}{W16}
\defcitealias{Wisnioski15}{W15}

\title[TFR]{The evolution of the Tully-Fisher relation between $z\sim2.3$ and $z\sim0.9$ with \kd}\thanks{Based on observations collected at the European Organisation for Astronomical Research in the Southern Hemisphere under ESO programs 092.A-0091, 093.A-0079, 094.A-0217, 095.A-0047, and 096.A-0025.}

\author{H.~\"Ubler\altaffilmark{1}, 
N.~M.~Förster~Schreiber\altaffilmark{1}, 
R.~Genzel\altaffilmark{1,}\altaffilmark{2},
E.~Wisnioski\altaffilmark{1}, 
S.~Wuyts\altaffilmark{3}, 
P.~Lang\altaffilmark{1,}\altaffilmark{4}, 
T.~Naab\altaffilmark{5}, 
A.~Burkert\altaffilmark{6,}\altaffilmark{1}, 
L.~J.~Tacconi\altaffilmark{1}, 
D.~J.~Wilman\altaffilmark{6,}\altaffilmark{1}, 
M.~Fossati\altaffilmark{6,}\altaffilmark{1}, 
J.~T.~Mendel\altaffilmark{1,}\altaffilmark{6}, 
A.~Beifiori\altaffilmark{6,}\altaffilmark{1}, 
S.~Belli\altaffilmark{1}, 
R.~Bender\altaffilmark{6,}\altaffilmark{1}, 
G.~B.~Brammer\altaffilmark{7}, 
J.~Chan\altaffilmark{6,}\altaffilmark{1}, 
R.~Davies\altaffilmark{1}, 
M.~Fabricius\altaffilmark{1}, 
A.~Galametz\altaffilmark{1,}\altaffilmark{6}, 
D.~Lutz\altaffilmark{1}, 
I.~G.~Momcheva\altaffilmark{7}, 
E.~J.~Nelson\altaffilmark{1}, 
R.~P.~Saglia\altaffilmark{1,}\altaffilmark{6}, 
S.~Seitz\altaffilmark{6,}\altaffilmark{1}, 
K.~Tadaki\altaffilmark{1}, 
P.~G.~van~Dokkum\altaffilmark{8} 
}
\altaffiltext{1}{Max-Planck-Institut f\"ur extraterrestrische Physik, Giessenbachstr.\ 1, D-85737 Garching, Germany; \href{mailto:hannah@mpe.mpg.de}{hannah@mpe.mpg.de}}
\altaffiltext{2}{Departments of Physics and Astronomy, University of California, Berkeley, CA 94720, USA}
\altaffiltext{3}{Department of Physics, University of Bath, Claverton Down, Bath, BA2 7AY, UK}
\altaffiltext{4}{Max-Planck-Institut f\"ur Astronomie, K\"onigstuhl 17, D-69117 Heidelberg, Germany}
\altaffiltext{5}{Max-Planck-Institut f\"ur Astrophysik, Karl Schwarzschildstr.\ 1, D-85737 Garching, Germany}
\altaffiltext{6}{Universit\"ats-Sternwarte Ludwig-Maximilians-Universit\"at M\"unchen, Scheinerstr.\ 1, D-81679 M\"unchen, Germany}
\altaffiltext{7}{Space Telescope Science Institute, 3700 San Martin Drive, Baltimore, MD 21218, USA}
\altaffiltext{8}{Department of Astronomy, Yale University, New Haven, CT 06511, USA}

\begin{abstract}
	We investigate the stellar mass and baryonic mass Tully-Fisher relations (TFRs) of massive star-forming disk galaxies at redshift $z\sim2.3$ and $z\sim0.9$ as part of the \kd integral field spectroscopy survey. 
	Our spatially resolved data allow reliable modelling of individual galaxies, including the effect of pressure support on the inferred gravitational potential. 
	At fixed circular velocity, we find higher baryonic masses and similar stellar masses at $z\sim2.3$ as compared to $z\sim0.9$.
	Together with the decreasing gas-to-stellar mass ratios with decreasing redshift, this implies that the contribution of dark matter to the dynamical mass on the galaxy scale increases towards lower redshift. 
	A comparison to local relations reveals a negative evolution of the stellar and baryonic TFR zero-points from $z=0$ to $z\sim0.9$, no evolution of the stellar TFR zero-point from $z\sim0.9$ to $z\sim2.3$, but a positive evolution of the baryonic TFR zero-point from $z\sim0.9$ to $z\sim2.3$.
	We discuss a toy model of disk galaxy evolution to explain the observed, non-monotonic TFR evolution, taking into account the empirically motivated redshift dependencies of galactic gas fractions, and of the relative amount of baryons to dark matter on the galaxy and halo scales.
\end{abstract}


\received{2016 December 23} \revised{2017 May 6} \accepted{2017 May 24}



\section{Introduction}\label{intro}

State-of-the-art cosmological simulations in a $\Lambda$CDM framework indicate that three main mechanisms regulate the growth of galaxies, namely the accretion of baryons, the conversion of gas into stars, and feedback. While gas settles down at the centers of growing dark matter (DM) haloes, cools and forms stars, it keeps in its angular momentum an imprint of the dark halo. 
Conservation of the net specific angular momentum, as suggested by analytical models of disk galaxy formation \citep[e.g.][]{Fall80, Dalcanton97, MMW98, Dutton07, Somerville08}, should result in a significant fraction of disk-like systems.
In fact, they make up a substantial fraction of the observed galaxy population at high redshift \citep[$1\lesssim z\lesssim 3$;][]{Labbe03, FS06b, FS09, Genzel06, Genzel14b, Law09, Epinat09, Epinat12, Jones10, Miller12, Wisnioski15, Stott16} and in the local Universe \citep[e.g.][and references therein]{Blanton09}. 
The detailed physical processes during baryon accretion from the halo scales to the galactic scales are, however, complex, and angular momentum conservation might not be straightforward to achieve \citep[e.g.][]{Danovich15}. 
To produce disk-like systems in numerical simulations, feedback from massive stars and/or active galactic nuclei is needed to prevent excessive star formation and to balance the angular momentum distribution of the star-forming gas phase \citep[e.g.][]{Governato07, Scannapieco09, Scannapieco12, Agertz11, Brook12a, Aumer13, Hopkins14, Marinacci14, Uebler14, Genel15}.
Despite the physical complexity and the diverse formation histories of individual galaxies, local disk galaxies exhibit on average a tight relationship between their rotation velocity $V$ and their luminosity $L$ or mass $M$, namely the Tully-Fisher relation \citep[TFR;][]{Tully77}. In its mass-based form, the TFR is commonly expressed as  $M\propto V^{a}$, or ${\rm log}(M)=a \cdot {\rm log}(V)+b$, where $a$ is the slope, and $b$ is the zero-point offset.

In the local Universe, rotation curves of disk galaxies are apparently generally dominated by DM already at a few times the disc scale length, and continue to be flat or rising out to several tens of kpc (see e.g.\ reviews by \citealp{Faber79, Sofue01}; and \citealp{Catinella06}). Therefore, the local TFR enables a unique approach to relate the baryonic galaxy mass, which is an observable once a mass-to-light conversion is assumed, to the potential of the dark halo.
Although the luminosity-based TFR is more directly accessible, relations based on mass constitute a physically more fundamental approach since the amount of light measured from the underlying stellar population is a function of passband, systematically affecting the slope of the TFR \citep[e.g.][]{Verheijen97, Verheijen01, Bell01, Pizagno07, Courteau07, McGaugh15}. The most fundamental relation is given by the baryonic mass TFR (bTFR). It places galaxies over several decades in mass onto a single relation, whereas there appears to be a break in the slope of the stellar mass TFR (sTFR) for low-mass galaxies \citep[][]{McGaugh00, McGaugh05}.

Observed slopes vary mostly between $3\lesssim a\lesssim4.5$ for the local sTFR \citep[e.g.][]{Bell01, Pizagno05, AvilaReese08, Williams10, Gurovich10, TorresFlores11, Reyes11} and between $3\lesssim a\lesssim4$ for the local bTFR \citep[e.g.][]{McGaugh00, McGaugh05, Trachternach09, Stark09, Zaritsky14, McGaugh15, Lelli16, Bradford16, Papastergis16}. 
It should be noted that the small scatter of local TFRs can be partly associated to the very efficient selection of undisturbed spiral galaxies (e.g.\ \citealp{Kannappan02}; see also \citealp{Courteau07, Lelli16}, for discussions of local TFR scatter).
Variations in the observational results of low-$z$ studies can be attributed to different sample sizes, selection bias, varying data quality, statistical methods, conversions from $L$ to $M$, or to the adopted measure of $V$ (\citealp{Courteau14}; for a detailed discussion regarding the bTFR see \citealp{Bradford16}). 

Any such discrepancy becomes more substantial when going to higher redshift where measurements are more challenging and the observed scatter of the TFR increases with respect to local relations \citep[e.g.][]{Conselice05, Miller12}. The latter is partly attributed to ongoing kinematic and morphological transitions \citep{Flores06, Kassin07, Kassin12, Puech08, Puech10, Covington10, Miller13, Simons16}, possibly indicating non-equilibrium states.
Another complication for comparing high-$z$ studies to local TFRs arises from the inherently different nature of the so-called disk galaxies at high redshift: although of disk-like structure and rotationally supported, they are significantly more ``turbulent'', geometrically thicker, and clumpier than local disk galaxies \citep[][]{FS06b, FS09, FS11a, FS11b, Genzel06, Genzel11, Elmegreen06, Elmegreen07, Kassin07, Kassin12, Epinat09, Epinat12, Law09, Law12, Jones10, Nelson12, Newman13, Wisnioski15, Tacchella15a, Tacchella15b}. 

Despite the advent of novel instrumentation and multiplexing capabilities, there is considerable tension in the literature regarding the empirical evolution of the TFR zero-points with cosmic time. 
Several authors find no or only weak zero-point evolution of the sTFR up to redshifts of $z\sim1.7$ \citep{Conselice05, Kassin07, Miller11, Miller12, Contini16, DiTeodoro16, Molina17, Pelliccia17}, while others find a negative zero-point evolution up to redshifts of $z\sim3$ \citep{Puech08, Puech10, Cresci09, Gnerucci11, Swinbank12, Price16, Tiley16, Straatman17}. 
Similarly for the less-studied high$-z$ bTFR, \cite{Puech10} find no indication of zero-point evolution since $z\sim0.6$, while \cite{Price16} find a positive evolution between lower-$z$ galaxies and their $z\sim2$ sample.
There are indications that varying strictness in morphological or kinematic selections can explain these conflicting results \citep{Miller13, Tiley16}. The work by \cite{Vergani12} demonstrates that also the assumed slope of the relation, which is usually adopted from a local TFR in high-$z$ studies, can become relevant for the debate of zero-point evolution \citep[see also][]{Straatman17}.

A common derivation of the measured quantities as well as similar statistical methods and sample selection are crucial to any study which aims at comparing different results and studying the TFR evolution with cosmic time \citep[e.g.][]{Courteau14, Bradford16}. 
Ideally, spatially well resolved rotation curves should be used which display a peak or flattening. Such a sample would provide an important reference frame for studying the effects of baryonic mass assembly on the morphology and rotational support of disk-like systems, for investigating the evolution of rotationally supported galaxies as a response to the structural growth of the parent DM halo, and for comparisons with cosmological models of galaxy evolution.

In this paper, we exploit spatially resolved integral field spectroscopic (IFS) observations of 240 rotation-dominated disk galaxies from the \kd survey \citep[][hereafter W15]{Wisnioski15} to study the evolution of the sTFR and bTFR between redshifts $z=2.6$ and $z=0.6$. The wide redshift coverage of the survey, together with its high quality data, allow for a unique investigation of the evolution of the TFR during the peak epoch of cosmic star formation rate density, where coherent data processing and analysis are ensured.
In Section~\ref{data} we describe our data and sample selection. We present the \kd TFR in Section~\ref{results}, together with a discussion of other selected high$-z$ TFRs. In Section~\ref{discussion} we discuss the observed TFR evolution, we set it in the context to local observations, and we discuss possible sources of uncertainties. In Section~\ref{conclusions} we constrain a theoretical toy model to place our observations in a cosmological context. Section~\ref{summary} summarizes our work.

Throughout, we adopt a \cite{Chabrier03} initial mass function (IMF) and a flat $\Lambda$CDM cosmology with $H_{0}=70 {\rm \,km\, s^{-1}\, Mpc^{-1}}$, $\Omega_{\Lambda}=0.7$, and $\Omega_{m}=0.3$.\\

\section{Data and sample selection}\label{data}

The contradictory findings about the evolution of the mass-based TFR in the literature motivate a careful sample selection at high redshift. 
In this work we concentrate on the evolution of the TFR for undisturbed disk galaxies.
Galaxies are eligible for such a study if the observed kinematics trace the central potential of the parent halo. To ensure a suitable sample we perform several selection steps which are described in the following paragraphs.

\subsection{The KMOS$^{3D}$ survey}\label{k3d}

This work is based on the first three years of observations of KMOS$^{\rm 3D}$, a multi-year near-infrared (near-IR) IFS survey of more than 600 mass-selected star-forming galaxies (SFGs) at $0.6\lesssim z\lesssim 2.6$ with the $K-$band Multi Object Spectrograph \citep[KMOS;][]{Sharples13} on the {\it Very Large Telescope}. The 24 integral field units of KMOS allow for efficient spatially resolved observations in the near-IR passbands $YJ$, $H$, and $K$, facilitating high-$z$ rest-frame emission line surveys of unprecedented sample size.
The \kd survey and data reduction are described in detail by \citetalias{Wisnioski15}, and we here summarize the key features. The \kd galaxies are selected from the 3D-HST survey, a {\it Hubble Space Telescope} Treasury Program \citep{Brammer12, Skelton14, Momcheva16}. 3D-HST provides $R\sim100$ near-IR grism spectra, optical to 8~$\mu$m photometric catalogues, and spectroscopic, grism, and/or photometric redshifts for all sources.
The redshift information is complemented by high-resolution Wide Field Camera 3 (WFC3) near-IR imaging from the CANDELS survey \citep{Grogin11, Koekemoer11, vdWel12}, as well as by further multi-wavelength coverage of our target fields GOODS-S, COSMOS, and UDS, through {\it Spitzer}/MIPS and {\it Herschel}/PACS photometry \citep[e.g.][and references therein]{Lutz11, Magnelli13, Whitaker14}.
Since we do not apply selection cuts other than a magnitude cut of $Ks\lesssim23$ and a stellar mass cut of log$(M_{*}~[M_{\odot}])\gtrsim9.2$, together with OH-avoidance around the survey's main target emission lines H$\alpha$+[N{\textsc{ii}}], the \kd sample will provide a reference for galaxy kinematics and H$\alpha$ properties of high$-z$ SFGs over a wide range in stellar mass and star formation rate (SFR).
The emphasis of the first observing periods has been on the more massive galaxies, as well as on $YJ-$ and $K-$band targets, i.e.\ galaxies at $z\sim0.9$ and $z\sim2.3$, respectively.
Deep average integration times of $5.5, 7.0, 10.5$~h in $YJ, H, K$, respectively, ensure a detection rate of more than 75 per cent, including also quiescent galaxies. 

The results presented in the remainder of this paper build on the \kd sample as of January 2016, with 536 observed galaxies. Of these, 316 are detected in, and have spatially resolved, H$\alpha$ emission free from skyline contamination, from which two-dimensional velocity and dispersion maps are produced. Examples of those are shown in the work by \citetalias{Wisnioski15} and \citet[][hereafter W16]{WuytsS16}.

\subsection{Masses and star-formation rates}\label{mass}

The derivation of stellar masses $M_{*}$ uses stellar population synthesis models by \cite{Bruzual03} to model the spectral energy distribution of each galaxy. Extinction, star formation histories (SFHs), and a fixed solar metallicity are incorporated into the models as described by \cite{WuytsS11a}. 

SFRs are obtained from the ladder of SFR indicators introduced by \cite{WuytsS11a}: if {\it Herschel}/PACS $60-160 \mu$m and/or {\it Spitzer}/MIPS $24 \mu$m observations were available, the SFRs were computed from the observed UV and IR luminosities. Otherwise, SFRs were derived from stellar population synthesis modelling of the observed broadband optical to IR spectral energy distributions.

Gas masses are obtained from the scaling relations by \cite{Tacconi17}, which use the combined data of molecular gas ($M_{\rm gas, mol}$) and dust-inferred gas masses of SFGs between $0<z<4$ to derive a relation for the depletion time $t_{\rm depl}\equiv M_{\rm gas, mol}/{\rm SFR}$. It is expressed as a function of redshift, main sequence offset, stellar mass, and size. 
Although the contribution of atomic gas to the baryonic mass within $1-3\, R_{e}$ is assumed to be negligible at $z\sim1-3$, the inferred gas masses correspond to lower limits \citep{Genzel15}.

Following \cite{Burkert16}, we adopt uncertainties of $0.15$~dex for stellar masses, and $0.20$~dex for gas masses. This translates into an average uncertainty of $\sim0.15$~dex for baryonic masses (see \S~\ref{mass_uncertainties} for a discussion).

\subsection{Dynamical modelling}\label{modelling}

\begin{figure}
	\centering
	\includegraphics[width=0.95\columnwidth]{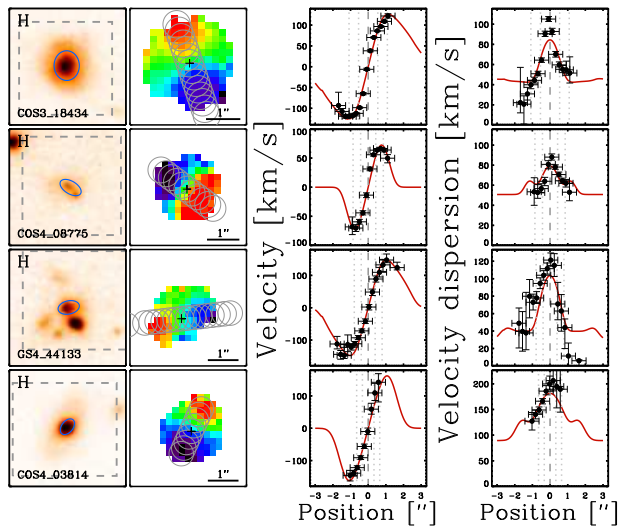}
    \caption{Examples of galaxies from the sample modelled by \citetalias{WuytsS16} which do, or do not, pass our TFR selection criteria (\S~\ref{tfr-sample}). From left to right: surface brightness distribution in the WFC3 $H-$band, with blue ellipses indicating the {\sc galfit} effective radius, and grey dashed lines marking the field of view of the KMOS observations; H$\alpha$ velocity field, with circles marking the extracted pseudo slit; the observed (black data points with errors) and modelled (red lines) 1D velocity and velocity dispersion profiles along the kinematic major axis, with vertical dotted grey lines marking one and two effective radii. More examples can be found in Figure~3 by \citetalias{WuytsS16}. 
    The upper two rows show galaxies which pass our selection criteria for the TFR sample. 
    The third row shows a galaxy which is rejected from the TFR sample because it is likely influenced by a neighboring object, based on projected distance, redshifts, and stellar mass ratio. 
    The bottom row shows a galaxy which is rejected from the TFR sample because it is unclear if the maximum velocity is covered by the observations.}
    \label{fig:examples}
\end{figure}

\citetalias{WuytsS16} use the two-dimensional velocity and velocity dispersion fields as observed in H$\alpha$ to construct dynamical models for selected galaxies. The modelling procedure is described in detail by \citetalias{WuytsS16}, where examples of velocity fields, velocity and dispersion profiles, and 1D fits can also be found (see also Figure~\ref{fig:examples}). 
In brief, radial velocity and dispersion profiles are constructed from $0\farcs8$ diameter circular apertures every other $0\farcs2$ along the kinematic major axis using {\sc linefit} \citep{Davies09}, where spectral resolution is taken into account. 
On average, the outermost apertures reach 2.5 times the effective $H$-band radius, corresponding to $\sim$15 and $\sim$12 extracted data points for galaxies at $z\sim0.9$ and $z\sim2.3$, respectively.
A dynamical mass modelling is performed by fitting the extracted kinematic profiles simultaneously in observed space using an updated version of {\sc dysmal} \citep{Cresci09, Davies11}. 

The free model parameters are the dynamical mass $M_{\rm dyn}$ and the intrinsic velocity dispersion $\sigma_{0}$.
The inclination $i$ and effective radius $R_{e}$ are independently constrained from {\sc galfit} \citep{Peng10} models to the CANDELS $H$-band imaging by {\it HST} presented by \cite{vdWel12}. The inclination is computed as ${\rm cos}(i)=[(q^2-q_0^2)/(1-q_0^2)]^{1/2}$. Here, $q=b/a$ is the axial ratio, and $q_0=0.25$ is the assumed ratio of scale height to scale length, representing the intrinsic thickness of the disk.
The width of the point spread function (PSF) is determined from the average PSF during observations for each galaxy. 
The mass model used in the fitting procedure is a thick exponential disk, following \cite{Noordermeer08}, with a Sérsic index of $n_{\rm S}=1$. 
We note that the peak rotation velocity of a thick exponential disk is about 3 to 8 per cent lower than that of a Freeman disk \citep{Freeman70}.
For a general comparison of observed and modelled rotation velocities and dispersions, we refer the reader to \citetalias{WuytsS16}. 
Another key product of the modelling is the baryonic (or DM) mass fraction on galactic scales, as presented in \citetalias{WuytsS16}.

The merit of the \citetalias{WuytsS16} modelling procedure includes the coupled treatment of velocity and velocity dispersion in terms of beam-smearing effects and pressure support. The latter is of particular importance for our study since high-$z$ galaxies have a non-negligible contribution to their dynamical support from turbulent motions \citep[][]{FS06b, FS09, Genzel06, Genzel08, Genzel14a, Kassin07, Kassin12, Cresci09, Law09, Gnerucci11, Epinat12, Swinbank12, Wisnioski12, Wisnioski15, Jones13, Newman13}. 
The resulting pressure compensates part of the gravitational force, leading to a circular velocity which is larger than the rotation velocity $v_{\rm rot}$ alone:
\begin{equation}\label{eq:vcirc}
	v_{\rm circ}(r)^{2}=v_{\rm rot}(r)^{2}+2\,\sigma_{0}^{2}\,\frac{r}{R_{d}},
\end{equation}
where $R_{d}$ is the disk scale length (\citealp{Burkert10}; see also \citealp{Burkert16}; \citealp{WuytsS16}; \citealp{Genzel17}; \citealp{Lang17}).

If not stated otherwise, we adopt the maximum of the modelled circular velocity, $v_{\rm{circ, max}}\equiv v_{\rm{circ}}$, as the rotation velocity measure for our Tully-Fisher analysis.
For associated uncertainties, see \S~\ref{vel_uncertainties}.
We use an expression for the peak velocity because there is strong evidence that high-$z$ rotation curves of massive star forming disk galaxies exhibit on average an outer fall-off, i.e.\ do not posses a `flat' part \citep{vDokkum15, Genzel17, Lang17}. This is partly due to the contribution from turbulent motions to the dynamical support of the disk, and partly due to baryons dominating the mass budget on the galaxy scale at high redshift \cite[see also][]{vDokkum15, Stott16, WuytsS16, Price16, Alcorn16, Pelliccia17}. A disk model with a flattening or rising rotation curve as the `arctan model', which is known to be an adequate model for local disk galaxies \citep[e.g.][]{Courteau97}, might therefore be a less appropriate choice for high-$z$ galaxies.

\subsection{Sample selection}\label{tfr-sample}

We start our investigation with a parent sample of 240 \kd galaxies selected and modelled by \citetalias{WuytsS16}.
The sample definition is described in detail by \citetalias{WuytsS16}, and we briefly summarize the main selection criteria here: 
(i) galaxies exhibit a continuous velocity gradient along a single axis, the `kinematic major axis'; 
(ii) their photometric major axis as determined from the CANDELS WFC3 $H$-band imaging and kinematic major axis are in agreement within 40 degrees; 
(iii) they have a signal-to-noise ratio within each $0\farcs8$ diameter aperture along the kinematic major axis of $S/N\gtrsim5$, with up to $S/N\sim10-100$ within the central apertures. 
The galaxies sample a parameter space along the main sequence of star forming galaxies (MS) with stellar masses of $M_{*}\gtrsim6.3\times10^{9}~M_{\odot}$, specific star formation rates of sSFR~$\gtrsim0.7/t_{\rm Hubble}$, and effective radii of $R_{e}\gtrsim2$~kpc.
The \citetalias{WuytsS16} sample further excludes galaxies with signs of major merger activity based on their morphology and/or kinematics.

For our Tully-Fisher analysis we undertake a further detailed examination of the \citetalias{WuytsS16} parent sample.
The primary selection step is based on the position-velocity diagrams and on the observed and modelled one-dimensional kinematic profiles of the galaxies. Through inspection of the diagrams and profiles we ensure that the peak rotation velocity is well constrained, based on the observed flattening or turnover in the rotation curve and the coincidence of the dispersion peak within $\lesssim2$ pixels ($\lesssim0\farcs4$) with the position of the steepest velocity gradient. 
The requirement of detecting the maximum velocity is the selection step with the largest effect on sample size, leaving us with 149 targets. The galaxy shown in the fourth row of Figure~\ref{fig:examples} is excluded from the TFR sample based on this latter requirement.

To single out rotation-dominated systems for our purpose, we next perform a cut of $v_{\rm rot, max}/\sigma_{0}>\sqrt{4.4}$, based on the properties of the modelled galaxy \citep[see also e.g.][]{Tiley16}.
Our cut removes ten more galaxies where the contribution of turbulent motions at the radius of maximum rotation velocity, which is approximately at $r=2.2\,R_{d}$, to the dynamical support is higher than the contribution from ordered rotation (cf. Equation~\eqref{eq:vcirc}).

We exclude four more galaxies with close neighbours because their kinematics might be influenced by the neighbouring objects. These objects have projected distances of $<20$~kpc, spectroscopic redshift separations of $<300$~km/s, and mass ratios of $>1:5$, based on the 3D-HST catalogue. One of the dismissed galaxies is shown in the third row of Figure~\ref{fig:examples}. 

After applying the above cuts, our refined TFR sample contains 135 galaxies, with 65, 24, 46 targets in the $YJ, H, K$ passbands with mean redshifts of $z\sim0.9, 1.5, 2.3$, respectively. 
The median and central 68\textsuperscript{th} percentile ranges of offsets between the morphological and kinematic position angle (PA) are $6.4^{\circ}$ $[0.1^{\circ};18.4^{\circ}]$. This should minimize the possible impact of non-axisymmetric morphological features on the fixed model parameters ($R_{e}$, sin($i$), PA) that are based on single-component Sérsic model fits to the observed $H$-band images (see \citealp{Rodrigues17}, and also the discussion by \citetalias{WuytsS16}).
The median physical properties of redshift subsamples are listed in Table~\ref{tab:properties}.
Individual properties of galaxies in the TFR sample in terms of $z$, $M_{*}$, $M_{\rm bar}$, $v_{\rm circ, max}$, and $\sigma_{0}$, are listed in Table~\ref{tab:ind-properties}.\\

\begin{figure}
\centering
 	\includegraphics[width=\columnwidth]{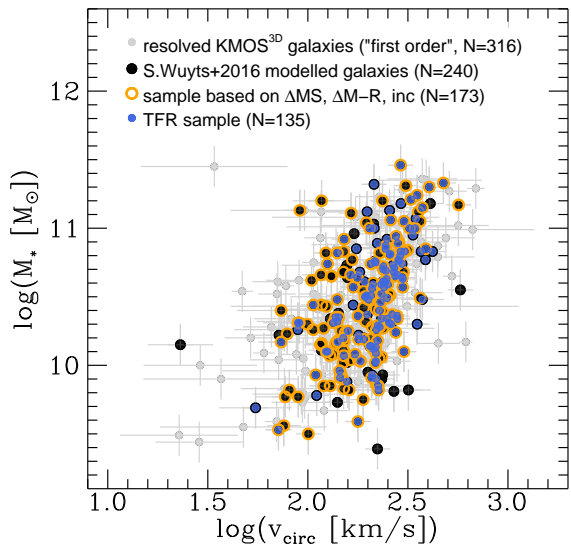}
   \caption{A `first order' sTFR of all detected and resolved \kd galaxies without skyline contamination at the position of H$\alpha$, where $v_{\rm circ}$ is computed from the observed maximal velocity difference and from the intrinsic velocity dispersion as measured from the outer disk region, after corrections for beam-smearing and inclination \citepalias[see][]{Wisnioski15}.
   The sample of galaxies which have been dynamically modelled by \citetalias{WuytsS16} is shown in black. In orange, we indicate a subsample of this latter sample based only on cuts in MS offset ($\pm0.6$~dex), mass-radius relation offset ($\pm0.3$~dex), and inclination ($0.5\leq {\rm sin}(i) \leq0.98$). In blue we show our final TFR sample as obtained from the selection steps outlined in \S~\ref{tfr-sample}.}
    \label{fig:selection}
\end{figure}

To visualize the impact of our sample selection we show in Figure~\ref{fig:selection} a `first order' sTFR of all detected and resolved \kd galaxies. Here, $v_{\rm circ}$ is computed from the observed maximal velocity difference and from the intrinsic velocity dispersion as measured from the outer disk region, after corrections for beam-smearing and inclination, as detailed in Appendix~A.2 of \cite{Burkert16}. For simplicity, we assume in computing $v_{\rm circ}$ for this figure that the observed maximal velocity difference is measured at $r=2.2R_{d}$, but we emphasize that, in contrast to the modelled circular velocity, this is not necessarily the case.
We indicate our parent sample of modelled galaxies by \citetalias{WuytsS16} in black, and our final TFR sample in blue. For reference, we also show in orange a subsample of the selection by \citetalias{WuytsS16} which is only based on cuts in MS offset ($\pm0.6$~dex), mass-radius relation offset ($\pm0.3$~dex), and inclination ($0.5\leq {\rm sin}(i) \leq0.98$). We emphasize that the assessment of recovering the true maximum rotation velocity is not taken into account for such an objectively selected sample.
We discuss in Appendix~\ref{selection-effects} in more detail the effects of sample selection, and contrast them to the impact of correcting for e.g.\ beam-smearing.

\begin{figure*}
\centering
	\includegraphics[width=0.45\textwidth]{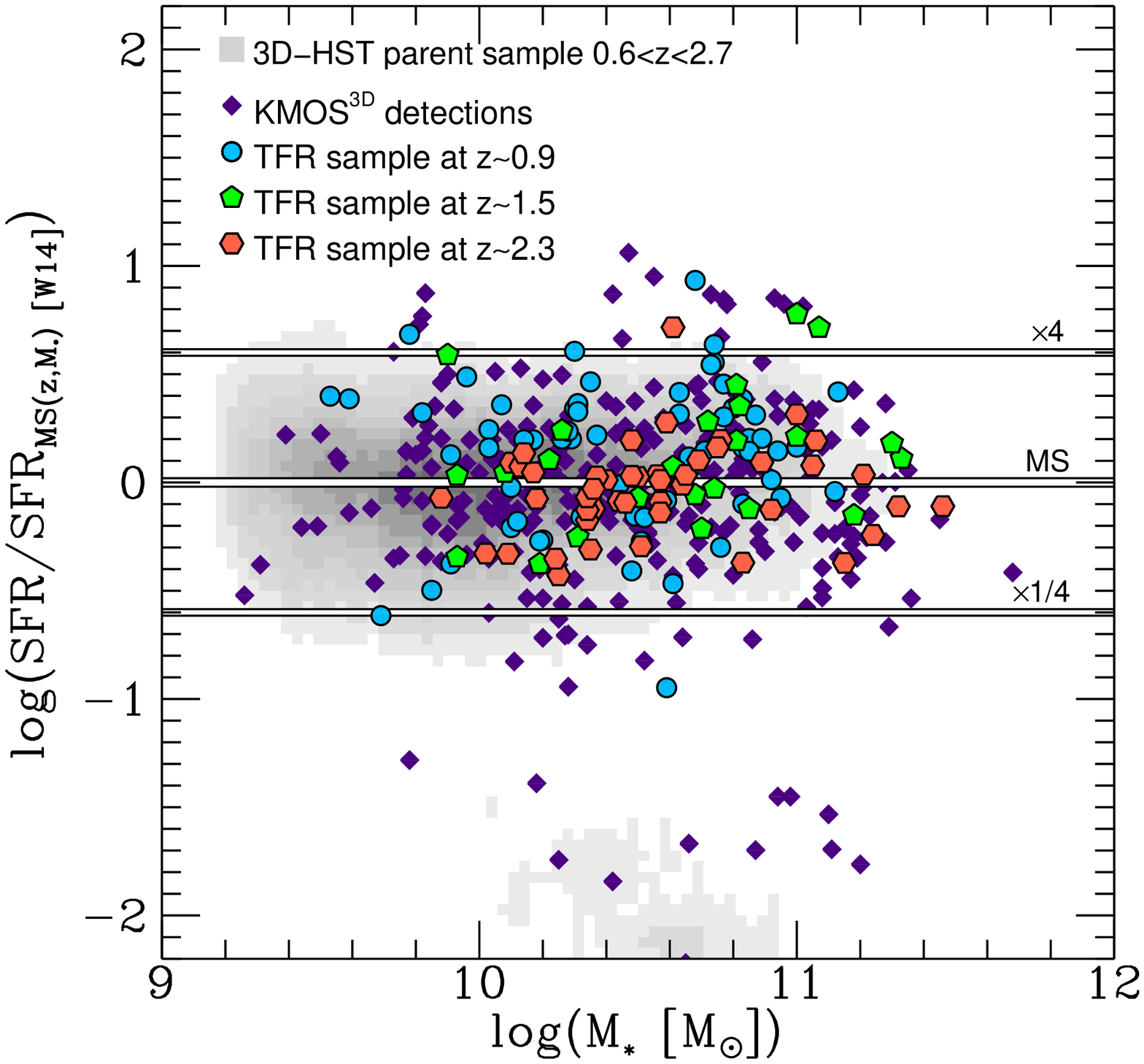}
	\hspace{8mm}
 	\includegraphics[width=0.45\textwidth]{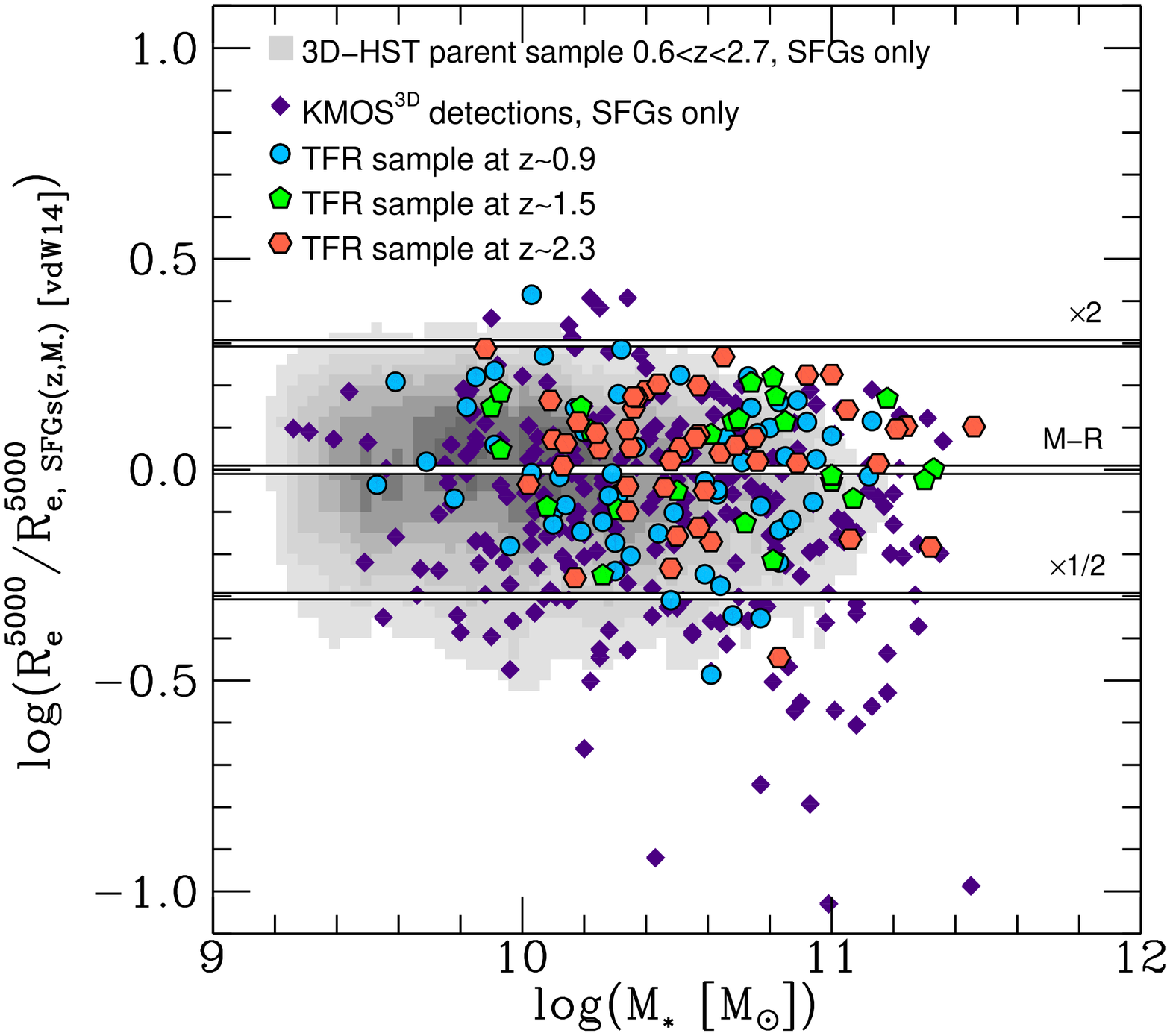}
   \caption{Location of our TFR galaxies in the $M_{*}$-SFR (left) and in the $M_{*}$-$R_{e}$ plane (right) as compared to all detected \kd galaxies (purple diamonds) and the underlying galaxy population at $0.6<z<2.7$ taken from the 3D-HST source catalogue (grey scale) with log$(M_{*}\, [M_{\odot}])>9.2$, $K_{\rm AB}<23$~mag, and for the $M_{*}$-$R_{e}$ relation sSFR~$>0.7/t_{\rm Hubble}$ (`SFGs only'). 
   In the left panel, the SFR is normalized to the MS as derived by \cite{Whitaker14} at the redshift and stellar mass of each galaxy, using the redshift-interpolated parametrization by \citetalias{Wisnioski15}. In the right panel, the effective radii as measured from $H-$band are corrected to the rest-frame $5000~\AA$ and normalized to the M-R relation of SFGs as derived by \cite{vdWel14a} at the redshift and stellar mass of each galaxy.
   At $z\sim0.9$ the TFR galaxies lie on average a factor of $\sim1.6$ above the MS, but on average on the M-R relation. At $z\sim2.3$, the TFR galaxies lie on average on the MS and the M-R relation, but their scatter with respect to higher SFRs and to smaller radii is not as pronounced as for the star-forming 3D-HST sample. For the 3D-HST `SFGs only' population the median and 68\textsuperscript{th} percentile ranges are log($\Delta$ MS)=$0.00^{+0.33}_{-0.37}$, and log($\Delta$ M-R)=$-0.04^{+0.17}_{-0.28}$. See Table~\ref{tab:properties} for the corresponding ranges of the TFR sample.}
    \label{fig:sfr-m}
\end{figure*}

The distribution of the TFR sample with respect to the full \kd sample (as of January 2016) and to the corresponding 3D-HST sample in terms of star formation rate and effective radius as a function of stellar mass is shown in Figure~\ref{fig:sfr-m} (for a detailed comparison of the \citetalias{WuytsS16} sample, we refer the reader to \citetalias{WuytsS16}). 
We select 3D-HST galaxies with $0.6<z<2.7$, log$(M_{*}~[M_{\odot}])>9.2$, $Ks<23$, and for the `SFGs only' subset we apply sSFR~$>0.7/t_{\rm Hubble}$, for a total of 9193 and 7185 galaxies, respectively. Focussing on the `SFGs only' subset, the median and corresponding 68\textsuperscript{th} percentiles with respect to the MS relations for the $z\sim0.9$ and the $z\sim2.3$ populations are log($\Delta$ MS)=$0.00^{+0.34}_{-0.39}$ and log($\Delta$ MS)=$-0.05^{+0.26}_{-0.35}$, and with respect to the mass-size (M-R) relation log($\Delta$ M-R)=$-0.04^{+0.16}_{-0.28}$ and log($\Delta$ M-R)=$-0.02^{+0.17}_{-0.31}$, respectively.
At $z\sim0.9$, the TFR galaxies lie on average a factor of $\sim1.6$ above the MS, with log($\Delta$ MS)=$0.20^{+0.42}_{-0.21}$, and have sizes corresponding to log($\Delta$ M-R)=$-0.02^{+0.16}_{-0.17}$. At $z\sim2.3$, the TFR galaxies lie on average on the MS and M-R relations (log($\Delta$ MS)=$-0.01^{+0.13}_{-0.29}$, log($\Delta$ M-R)=$0.06^{+0.17}_{-0.14}$), but their scatter with respect to higher SFRs and to smaller radii is not as pronounced as for the star-forming 3D-HST sample.

\begin{table*}
\centering
 \caption{Median physical properties of our TFR subsamples at $z\sim0.9$ ($YJ$), $z\sim1.5$ ($H$), and $z\sim2.3$ ($K$), together with the associated central 68\textsuperscript{th} percentile ranges in brackets.}
 \label{tab:properties}
 \begin{tabular}{lrrr}
  \tableline
  & $z\sim0.9$ & $z\sim1.5$ & $z\sim2.3$\\
  & (65 galaxies) & (24 galaxies) & (46 galaxies)\\
  \tableline
  log(M$_{*}$ [M$_{\odot}$]) & 10.49 [10.03; 10.83] & 10.72 [10.08; 11.07] & 10.51 [10.18; 11.00] \\
  log(M$_{\rm bar}$ [M$_{\odot}$]) & 10.62 [10.29; 10.98] & 10.97 [10.42; 11.31] & 10.89 [10.59; 11.33] \\
  SFR [M$_{\odot}$/yr] & 21.1 [7.1; 39.6] & 53.4 [15.5; 134.5] & 72.9 [38.9; 179.1] \\
  log($\Delta$ MS)\tablenotemark{a} & 0.20 [-0.21; 0.42] & 0.10 [-0.21; 0.45] & -0.01 [-0.29; 0.13] \\
  $R_{e}^{5000}$ [kpc] & 4.8 [3.0; 7.6] & 4.9 [3.0; 7.0] & 4.0 [2.5; 5.2] \\
  log($\Delta$ M-R)\tablenotemark{b} & -0.02 [-0.17; 0.16] & 0.08 [-0.10; 0.17] & 0.06 [-0.14; 0.17] \\
  $n_{\rm S}$ & 1.3 [0.8; 3.1] & 0.9 [0.4; 2.2] & 1.0 [0.4; 1.6] \\
  $B/T$\tablenotemark{c} & 0.11 [0.00; 0.39] & 0.00 [0.00; 0.23] & 0.10 [0.00; 0.25] \\
  $v_{\rm rot, max}$ [km/s] & 233 [141; 302] & 245 [164; 337] & 239 [160; 284] \\
  $\sigma_{0}$ [km/s] & 30 [9; 52] & 47 [29; 59] & 49 [32; 68] \\
  $v_{\rm rot, max}/\sigma_{0}$ & 6.7 [3.2; 25.3] & 5.5 [3.4; 65.6] & 4.3 [3.4; 9.1] \\
  $v_{\rm circ, max}$ [km/s] & 239 [167; 314] & 263 [181; 348] & 260 [175; 315] \\
  \tableline
 \end{tabular}
 \tablenotetext{a}{MS offset with respect to the broken power law relations derived by \cite{Whitaker14}, using the redshift-interpolated parametrization by \citetalias{Wisnioski15}, $\Delta$~MS=${\rm SFR - SFR_{MS(z,M_{*}) [W14]}}$.}
 \tablenotetext{b}{Offset from the mass-size relation of SFGs with respect to the relation derived by \cite{vdWel14a}, $\Delta$~M-R=$R_{e}^{5000} - R^{5000}_{e, {\rm M-R(z,M_{*}) [vdW14]}}$, after correcting the $H-$band $R_{e}$ to the rest-frame $5000~\AA$.}
 \tablenotetext{c}{Bulge-to-total mass ratio if available, namely for 78, 92, and 89 per cent of our galaxies in $YJ-$, $H-$, and $K-$band, respectively. Values of $B/T=0$ usually occur when the galaxy's Sérsic index $n_{\rm S}$ is smaller than 1 \citep[cf.][]{Lang14}.}
\end{table*}

In summary, our analysis accounts for the following effects: (i) beam-smearing, through a full forward modelling of the observed velocity and velocity dispersion profiles with the known instrumental PSF; (ii) the intrinsic thickness of high$-z$ disks, following \cite{Noordermeer08}; (iii) pressure support through turbulent gas motions, following \cite{Burkert10}, under the assumption of a disk of constant velocity dispersion and scale height. The former steps are all included in the dynamical modelling by \citetalias{WuytsS16}. On top of that, we retain in our TFR sample only non-interacting SFGs which are rotationally supported based on the $v_{\rm rot,max}/\sigma_{0}>\sqrt{4.4}$ criterion, and for which the data have sufficient $S/N$ and spatial coverage to robustly map, and model, the observed rotation curve to or beyond the peak rotation velocity.\\

\section{The TFR with \kd}\label{results}

\subsection{Fitting}\label{fitting}

In general, there are two free parameters for TFR fits in log-log space: the slope $a$ and the zero-point offset $b$. It is standard procedure to adopt a local slope for high$-z$ TFR fits\footnote{While the slope might in principle vary with cosmic time, a redshift evolution is not expected from the toy model introduced in Section~\ref{intro}.}. This is due to the typically limited dynamical range probed by the samples at high redshift which makes it challenging to robustly constrain $a$.
The TFR evolution is then measured as the relative difference in zero-point offsets \citep[e.g.][]{Puech08, Cresci09, Gnerucci11, Miller11, Miller12, Tiley16}. In Appendix~\ref{alter-offset} we briefly investigate a method to measure TFR evolution which is independent of the slope. For clarity and consistency with TFR investigations in the literature, however, we present our main results based on the functional form of the TFR as given in Equation~\eqref{eq:fit} below. For our fiducial fits, we adopt the local slopes by \cite{Reyes11} and \cite{Lelli16} for the sTFR and the bTFR, respectively.\footnote{The sTFR zero-point by \citet{Reyes11} is corrected by $-0.034$~dex to convert their \cite{Kroupa01} IMF to the Chabrier IMF which is used in this work, following the conversions given in \cite{Madau14}.} 

To fit the TFR we adopt an inverse linear regression model of the form
\begin{equation}\label{eq:fit}
	{\rm log}(M~[M_{\odot}])=a \cdot {\rm log}(v_{\rm circ}/v_{\rm ref}) + b.
\end{equation}
Here, $M$ is the stellar or baryonic mass, and a reference value of $v_{\rm ref}=\overline{v_{\rm circ}}$ is chosen to minimize the uncertainty in the determination of the zero-point~$b$ \citep[][]{Tremaine02}. If we refer in the remainder of the paper to $b$ as the zero-point offset, this is for our sample in reference to $v_{\rm circ}=v_{\rm ref}$, and not to log($v_{\rm circ}$ [km/s])=0. When comparing to other data sets in \S\S~\ref{comp_high} and \ref{comp_local} we convert their zero-points accordingly.

For the fitting we use a Bayesian approach to linear regression, as well as a least-squares approximation. 
The Bayesian approach to linear regression takes uncertainties in ordinate and abscissa into account.\footnote{
We use the IDL routine {\sc linmix\_err} which is described and provided by \cite{Kelly07}. A modified version of this code which allows for fixing of the slope was kindly provided to us by Brandon Kelly and Marianne Vestergaard.} 
The least-squares approximation also takes uncertainties in ordinate and abscissa into account, and allows for an adjustment of the intrinsic scatter to ensure for a goodness of fit of $\chi^{2}_{\rm reduced}\approx1$.\footnote{
We use the IDL routine {\sc mpfitexy} which is described and provided by \cite{Williams10}. It depends on the {\sc mpfit} package \citep{Markwardt09}. }
To evaluate the uncertainties of the zero-point offset $b$ of the fixed-slope fits, a bootstrap analysis is performed for the fits using the least-squares approximation. The resulting errors agree with the error estimates from the Bayesian approach within $0.005$~dex of mass. 
We find that the intrinsic scatter obtained from the Bayesian technique is similar or larger by up to $0.03$~dex of mass as compared to the least-squares method. 
Both methods give the same results for the zero-point $b$ \citep[see also the recent comparison by][]{Bradford16}. 

We perform fits to our full TFR sample, as well as to the subsets at $z\sim0.9$ and $z\sim2.3$. The latter allows us to probe the maximum separation in redshift possible within the \kd survey. Due to the low number of TFR galaxies in our $H-$band bin we do not attempt to fit a zero-point at $z\sim1.5$.

\subsection{The TFR at $0.6<z<2.6$}\label{tfr}

\begin{figure*}
	\centering
	\includegraphics[width=0.45\textwidth]{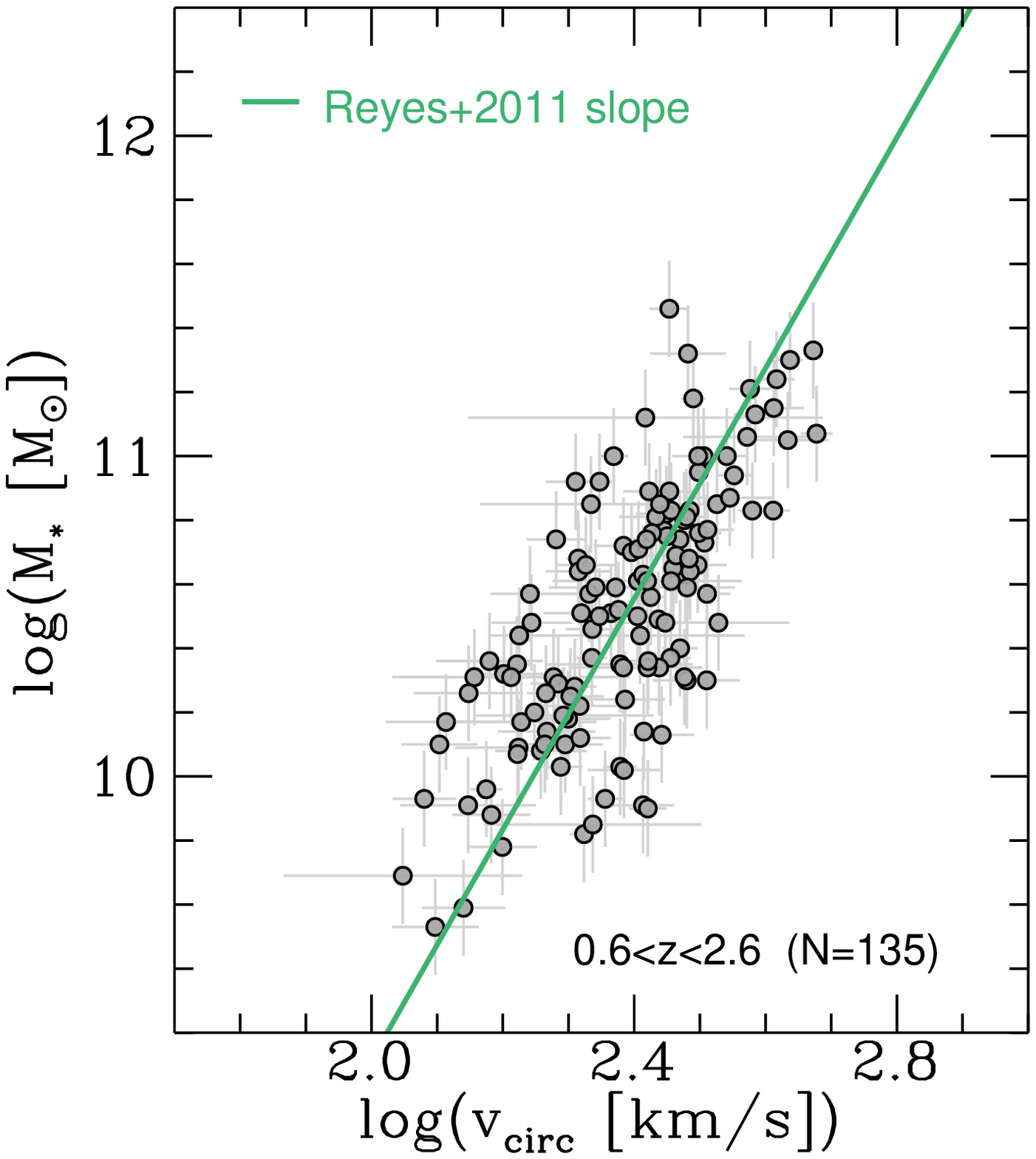}
	\hspace{8mm}
	\includegraphics[width=0.45\textwidth]{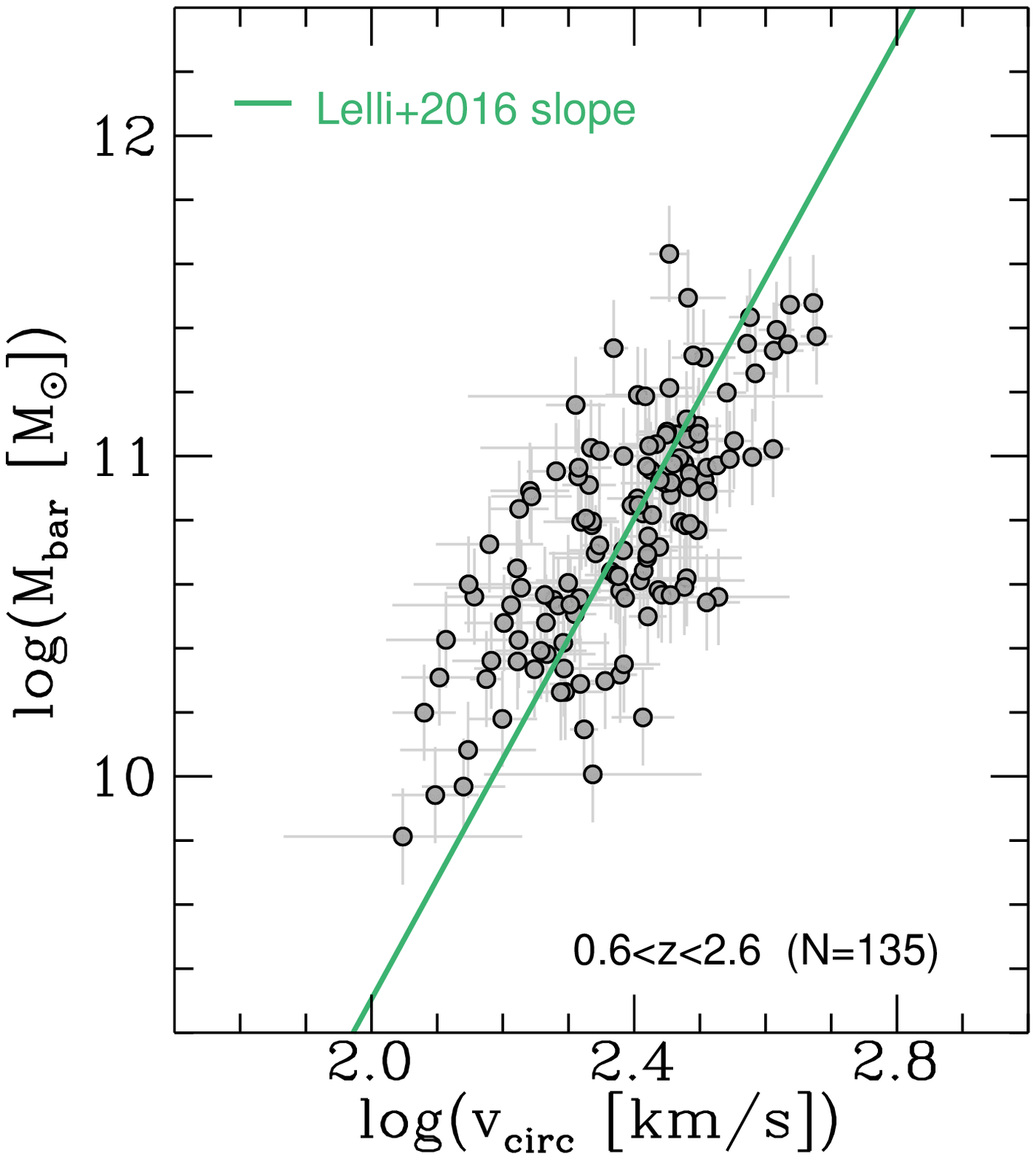}
    \caption{The sTFR (left) and the bTFR (right) for our sample of 135 SFGs, with error bars in grey. The green lines show the fixed-slope fits to the inverse linear regression model as given in Equation~\eqref{eq:fit}, using the corresponding local slopes by \cite{Reyes11} and \cite{Lelli16}. The fit parameters are given in Table~\ref{tab:fits}. A correlation between $v_{\rm circ}$ and the different mass tracers is evident.}
    \label{fig:free-fit}
\end{figure*}

In this paragraph, we investigate the Tully-Fisher properties of our full TFR sample at $0.6<z<2.6$. 
The sTFR as well as the bTFR are clearly in place and well defined at $0.6<z<2.6$, confirming previous studies \citep[e.g.][and other high$-z$ work cited in Section~\ref{intro}]{Cresci09, Miller11, Miller12, Tiley16}.
In Figure~\ref{fig:free-fit} we show the best fits for the sTFR and the bTFR using the local slopes by \cite{Reyes11} ($a=1/0.278=3.60$) and \cite{Lelli16} ($a=3.75$), respectively. 
The best-fit parameters are given in Table~\ref{tab:fits}. 

\begin{table*}
\centering
 \caption{Results from the inverse linear regression fits to Equation~\eqref{eq:fit} using the least-squares method, including bootstrapped errors of the zero-point. The reference velocity is $v_{\rm ref}=242$~km/s.}
 \label{tab:fits}
 \begin{tabular}{lcclll}
  \tableline
  TFR & redshift range & number of galaxies & slope $a$ (local relation) & zero-point $b$ (error) & intrinsic scatter $\zeta_{\rm int}$\\ 
  	& & & $\left[\frac{{\rm log}(M~[M_{\odot}])}{{\rm log}(v_{\rm circ}~[{\rm km/s}])}\right]$ & [log($M~[M_{\odot}$])] & [dex of $M_{\odot}$] \\
  \tableline
  sTFR & $0.6<z<2.6$ & 135 & 3.60 \citep{Reyes11} & 10.50 ($\pm0.03$) & 0.22 \\
  & $z\sim0.9$ & 65 & 3.60 \citep{Reyes11} & 10.49 ($\pm0.04$) & 0.21 \\
  & $z\sim2.3$ & 46 & 3.60 \citep{Reyes11} & 10.51 ($\pm0.05$) & 0.26 \\
 bTFR  & $0.6<z<2.6$ & 135 & 3.75 \citep{Lelli16} & 10.75 ($\pm0.03$) & 0.23 \\
  & $z\sim0.9$ & 65 & 3.75 \citep{Lelli16} & 10.68 ($\pm0.04$) & 0.22 \\
  & $z\sim2.3$ & 46 & 3.75 \citep{Lelli16} & 10.85 ($\pm0.05$) & 0.26 \\
  \tableline
 \end{tabular}
\end{table*}

The intrinsic scatter as determined from the fits is with $\zeta_{\rm int, sTFR}\approx0.22$ and $\zeta_{\rm int, bTFR}\approx0.23$ larger by up to a factor of two in dex of mass than in the local Universe (typical values for the observed intrinsic scatter of the local relations used in this study are $\zeta_{\rm int}=0.1-0.13$ in dex of mass; see \citealp{Reyes11, Lelli16}). 
A larger scatter in the high$-z$ TFR is expected simply due to the larger measurement uncertainties. It might further be due to disk galaxies being less ``settled'' (\citealp{Kassin07, Kassin12, Simons16}; see also \citealp{Flores06, Puech08, Puech10, Covington10, Miller13}). This can become manifest through actual displacement of galaxies from the TFR due to a non-equilibrium state (see e.g.\ simulations by \citealp{Covington10}).

\cite{Miller13} studied the connection between TFR scatter and bulge-to-total ratio, and found that above $z\approx1$ the TFR scatter is increased due to an offset of bulge-less galaxies from the $B/T>0.1$ galaxy population. 
$B/T$ measurements for our galaxies come from bulge-disk decompositions based on two-component fits to the two-dimensional CANDELS $H$-band light distribution \citep{Lang14}. 
If we select only galaxies with $B/T>0.1$ (57 galaxies), we do not find a decrease in scatter for our sample ($\zeta_{{\rm int, sTFR,} B/T>0.1}=0.22$ and $\zeta_{{\rm int, bTFR,} B/T>0.1}=0.24$). The same is true if we select for galaxies with $B/T<0.1$ (78 galaxies), leading to $\zeta_{{\rm int, sTFR,} B/T<0.1}=0.23$ and $\zeta_{{\rm int, bTFR,} B/T<0.1}=0.22$.

However, the scatter is affected by the sample selection: if we create `first order' TFRs (\S~\ref{tfr-sample}, Figure~\ref{fig:selection}), i.e.\ using all detected and resolved \kd galaxies without skyline contamination (316 SFGs), but also without selecting against dispersion-dominated systems, low $S/N$ galaxies, or mergers, we find an intrinsic scatter of $\zeta_{\rm int,sTFR}=0.60$ and $\zeta_{\rm int,bTFR}=0.64$ for these `first order' TFRs (for the parent \citetalias{WuytsS16} sample we find $\zeta_{\rm int,sTFR}=0.27$ and $\zeta_{\rm int,bTFR}=0.29$). We caution that this test sample includes galaxies where the maximum rotation velocity is not reached, thus introducing artificial scatter in these `first order' TFRs. In contrast to the properties of our TFR sample, this scatter is asymmetric around the best fit, with larger scatter towards lower velocities, but also towards lower masses where more of the dispersion-dominated galaxies reside (cf.\ Figures~\ref{fig:selection} and \ref{fig:vcorr}).
This underlines the importance of a careful sample selection. 

Also the zero-points are affected by the sample selection (see also Figure~\ref{fig:vcorr}). For our TFR sample, we find $b_{\rm sTFR}=10.50\pm0.03$ and $b_{\rm bTFR}=10.75\pm0.03$. If we consider the `first order' samples we find an increase of the zero-points of $\Delta b_{\rm sTFR}=0.37$~dex and $\Delta b_{\rm bTFR}=0.39$~dex (for the parent \citetalias{WuytsS16} sample we find $\Delta b_{\rm sTFR}=0.03$~dex and $\Delta b_{\rm bTFR}=0.04$~dex).

It is common, and motivated by the scatter of the TFR, to investigate the existence of hidden parameters in the relation. 
For example, a measure of the galactic radius (effective, or exponential scale length) has been investigated by some authors to test for correlations with TFR residuals \citep[e.g.][]{McGaugh05, Pizagno05, Gnedin07, Zaritsky14, Lelli16}. The radius, together with mass, determines the rotation curve (e.g.\ Equation~\eqref{eq:freeman}). 
Adopting the local slopes, we do not find significant correlations (based on Spearman tests) of the TFR residuals with $R_{e}$, $B/T$, $n_{\rm S}$, stellar or baryonic mass surface density, offset from the main sequence or the mass-radius relation, SFR surface density $\Sigma_{\rm SFR}$, or inclination. 
In Appendix~\ref{uncertainties} we investigate how the uncertainties in stellar and baryonic mass affect second-order parameter dependencies for TFR fits with free slopes, by example of $R_{e}$ and $\Sigma_{\rm SFR}$.

In summary, we find well defined mass-based TFRs at $0.6<z<2.6$ for our sample. If galaxies with underestimated peak velocity, dispersion-dominated and disturbed galaxies are included, the TFR zero-points are increasing, and also the scatter increases, especially towards lower velocities and masses. Adopting the local slopes, we find no correlation of TFR residuals with independent galaxy properties.

\subsection{TFR evolution from $z\sim2.3$ to $z\sim0.9$}\label{zp-ev}

\begin{figure*}
	\centering
 	\includegraphics[width=\columnwidth]{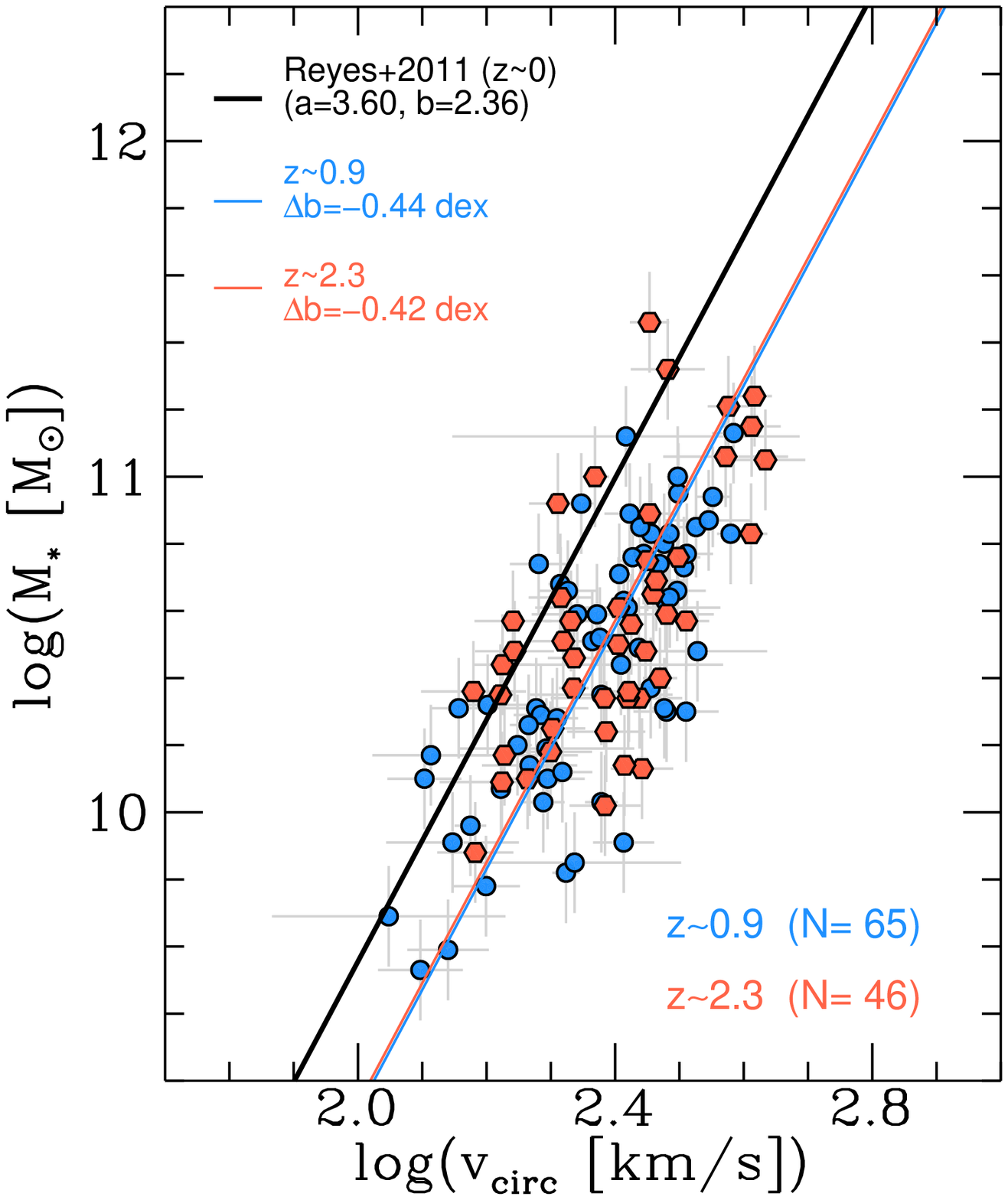}
	\hspace{8mm}
	\includegraphics[width=\columnwidth]{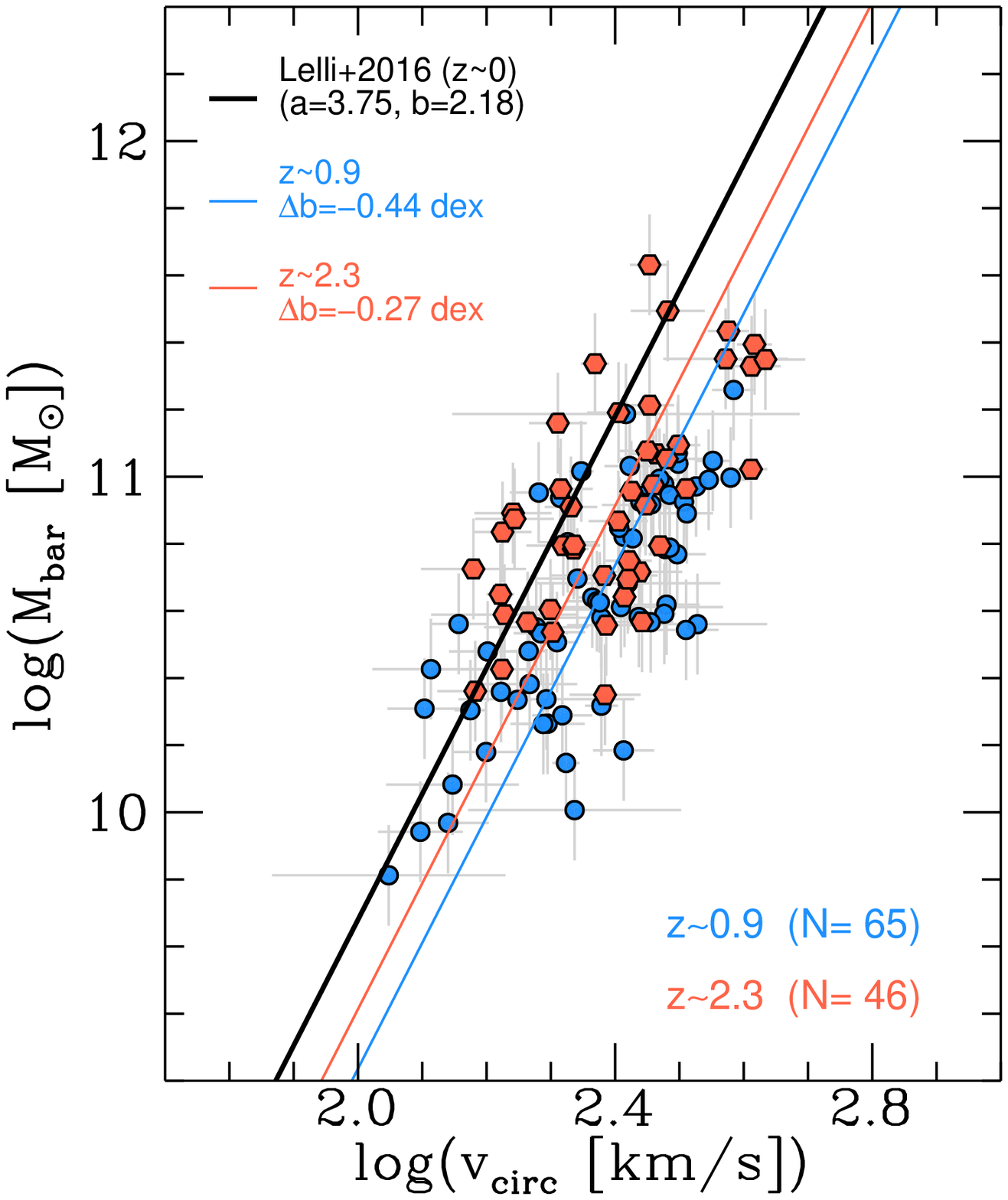}
    \caption{Fixed-slope fits for the sTFR (left) and the bTFR (right) using local (black) slopes to our \kd subsamples at $z\sim0.9$ (blue) and $z\sim2.3$ (red). For the local relations, we give $a$ and $b$ corresponding to our adopted functional form of the TFR give in Equation~\eqref{eq:fit}, with log($v_{\rm ref}$~[km/s])=0.
    For the sTFR, we find no (or only marginal) evolution of the sTFR zero-point in the studied redshift range. Comparing to the local relation by \cite{Reyes11} we find $\Delta b=-0.44$ and $-0.42$~dex at $z\sim0.9$ and $z\sim2.3$, respectively.
    For the bTFR, we find a positive evolution of the zero-point between $z\sim0.9$ and $z\sim2.3$. 
    Comparing to the local relation by \cite{Lelli16} we find  $\Delta b=-0.44$ and $-0.27$~dex at $z\sim0.9$ and $z\sim2.3$, respectively.}
    \label{fig:stfr_btfr}
\end{figure*}

We now turn to the TFR subsamples at $z\sim0.9$ and $z\sim2.3$. We adopt the local slopes by \cite{Reyes11} and \cite{Lelli16} to investigate the zero-point evolution.
Our redshift subsamples are shown in Figure~\ref{fig:stfr_btfr} for the sTFR (left) and bTFR (right), together with the corresponding local relations and the respective fixed-slope fits. The parameters of each fit are given in Table~\ref{tab:fits}.

For the sTFR we find no indication for a significant change in zero-point between $z\sim0.9$ and $z\sim2.3$ within the best fit uncertainties. Using the local slope of $a=3.60$ and the reference value $v_{\rm ref}=242$~km/s, we find a zero-point of $b=10.49\pm0.04$ for the subsample at $z\sim0.9$, and of $b=10.51\pm0.05$ for the subsample at $z\sim2.3$, translating into a zero-point evolution of $\Delta b=0.02$~dex between $z\sim0.9$ and $z\sim2.3$.

For the bTFR, however, using the local slope of $a=3.75$, and again the reference value $v_{\rm ref}=242$~km/s, we find a positive zero-point evolution between $z\sim0.9$ and $z\sim2.3$, with $b=10.68\pm0.04$ and $b=10.85\pm0.05$, respectively, translating into a zero-point evolution of $\Delta b=0.17$~dex between $z\sim0.9$ and $z\sim2.3$.

If we consider the `first order' TFR subsamples at $z\sim0.9$ and $z\sim2.3$, we find significantly different zero-point evolutions of $\Delta b_{\rm sTFR}=0.23$~dex and $\Delta b_{\rm bTFR}=0.28$~dex between $z\sim0.9$ and $z\sim2.3$. Again, this highlights the importance of a careful sample selection for TFR studies. Figure~\ref{fig:w16_tfr} shows that if instead we extend our data set to the sample from \citetalias{WuytsS16}, we find qualitatively the same trends as for the adopted TFR sample, namely an evolution of $\Delta b_{\rm sTFR}=0.05$~dex and $\Delta b_{\rm bTFR}=0.20$~dex for the zero-point between $z\sim0.9$ and $z\sim2.3$ (see Appendix~\ref{selection-effects}). 
Also, if we consider only TFR galaxies with $B/T>0.1(<0.1)$, our qualitative results remain the same.

In summary, we find no evolution for the sTFR, but a positive evolution of the bTFR between $z\sim0.9$ and $z\sim2.3$. If galaxies with underestimated peak velocity, dispersion-dominated and disturbed galaxies are included, we find positive evolution of both the sTFR and the bTFR.

\subsection{Comparison to other high$-z$ studies}\label{comp_high}

\begin{figure*}
	\includegraphics[width=\textwidth]{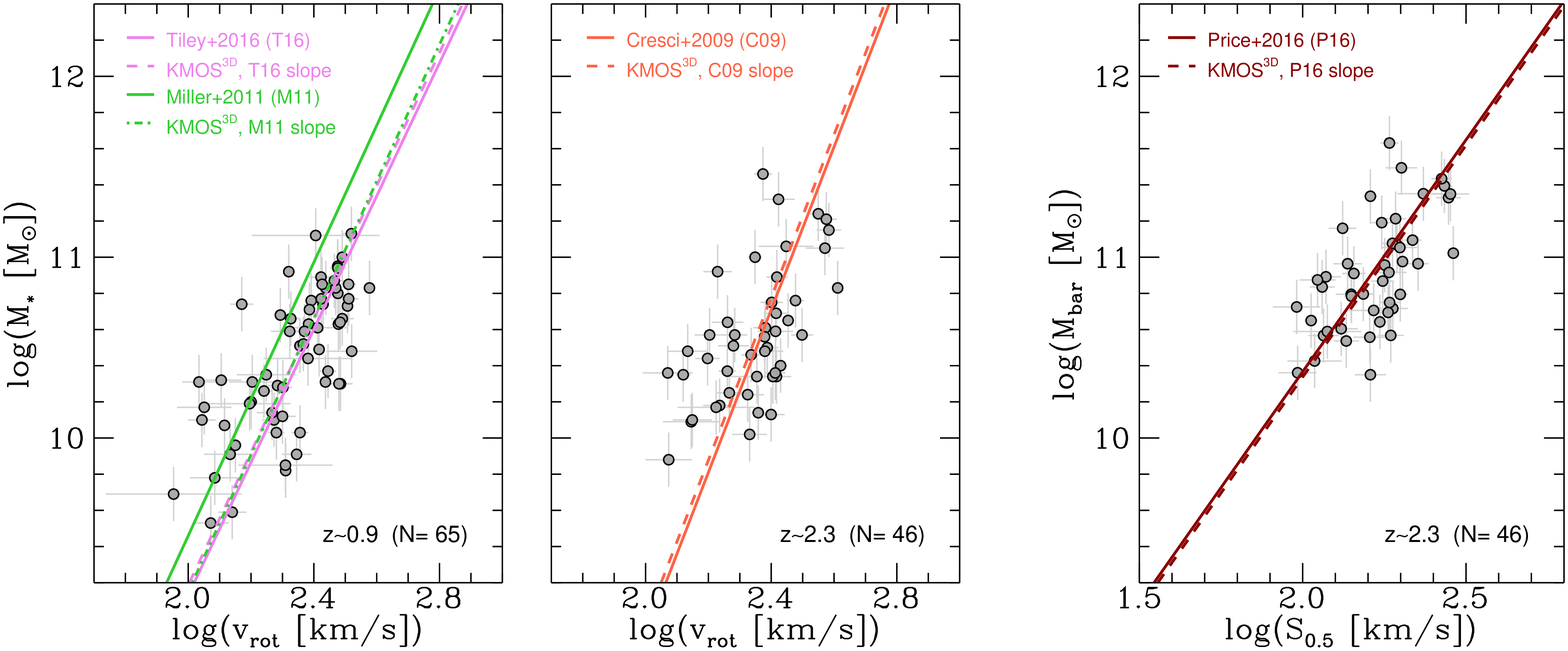}
    \caption{{\bf Left and middle panel:} the $v_{\rm rot}$-sTFRs at $z\sim0.9$ (left panel) and $z\sim2.3$ (middle panel). We show fits from \cite{Tiley16} ($z\sim0.9$; magenta), \cite{Miller11} ($z\sim1$; green) and \cite{Cresci09} ($z\sim2.2$; orange) as solid lines, together with corresponding fixed-slope fits to our samples as dashed lines. From \citet{Tiley16}, we use their best fixed-slope fit to their disky subsample. From \citet{Miller11}, we use the $z\sim1$ fit corresponding to total stellar mass and $v_{\rm rot,3.2}$. Our findings regarding the zero-point offset are in agreement with \citet{Tiley16} and \citet{Cresci09}, but in disagreement with \citet{Miller11}.
    {\bf Right panel:} the S$_{0.5}$-bTFR at $z\sim2.3$. We show the fit from \cite{Price16} ($z\sim2$; red) as a solid line, together with the corresponding fixed-slope fit to our sample as a dashed line. Our findings regarding the zero-point offset are in agreement.}
    \label{fig:stfr_lit}
\end{figure*}

At $z\sim0.9$ we compare our sTFR (65 \kd galaxies) to the work by \cite{Tiley16} and \cite{Miller11}. \cite{Tiley16} have investigated the sTFR at $z\sim0.9$ using 56 galaxies from the KROSS survey with KMOS \citep{Stott16}. 
\cite{Miller11, Miller12} have presented an extensive slit-based sTFR study at $0.2<z<1.7$ with 37 galaxies at $z\sim1$. From \citet{Tiley16}, we use their best fixed-slope fit to their disky subsample ($a=3.68$). From \citet{Miller11}, we use the $z\sim1$ fit corresponding to total stellar mass and $v_{\rm rot,3.2}$ ($a=3.78$).
For a sTFR comparison at $z\sim2.3$ (46 \kd galaxies), we consider the work by \cite{Cresci09}. The authors have studied the sTFR at $z\sim2.2$ for 14 galaxies from the SINS survey ($a=4.5$). Despite the small sample size, the high-quality data based on the 2D modelling of velocity and velocity dispersion maps qualify the sample for comparison with our findings in the highest redshift bin.

In the following, we use $v_{\rm rot,max}$ to ensure a consistent comparison with the measurements presented in these studies. 
For a comparison with the literature data, we make the simplifying assumption that $v_{\rm rot,max}$ is comparable to $v_{\rm rot,80}$ and $v_{\rm rot,3.2}$ (see \S~\ref{vel_measures} for a discussion).
We adopt the slopes reported in the selected studies to guarantee consistency in the determination of zero-point offsets. The results are shown in Figure~\ref{fig:stfr_lit} as dashed lines, while the original relations from the literature are shown as solid lines. 
The difference in zero-points, $\Delta b$, is then computed as the zero-point from the \kd fixed-slope fit minus the zero-point from the literature. Given the typical zero-point uncertainty of our fits of $\delta b\approx0.05$~dex, our results are in agreement with \citet{Tiley16} ($\Delta b=0.06$) and \citet{Cresci09} ($\Delta b=0.07$), but in disagreement with \citet{Miller11} at $z\sim1$ ($\Delta b=-0.31$). 
We further note that our findings are in disagreement with the recent study by \cite{DiTeodoro16} who employed a tilted ring model on a small subset of galaxies from the \kd and KROSS surveys at $z\sim1$ ($\Delta b=-0.34$; see also \citealp{Tiley16}).

A number of complications might give rise to conflicting results of different TFR studies, such as the use of various kinematic models, velocity tracers, mass estimates, or statistical methods.
\cite{Tiley16}, who present an extensive comparison of several sTFR studies from the literature, argue that conflicting results regarding the zero-point evolution with redshift depend on the ability of the studies to select for rotationally supported systems. 
The two-dimensional information on the velocity and velocity dispersion fields is a major advantage of IFS observations as it allows for the robust determination of the kinematic center and major axis.

We test the case of selecting against dispersion-dominated or disturbed systems for our TFR samples. For the full sample of 240 SFGs by \citetalias{WuytsS16}, which includes some dispersion-dominated systems and cases where the peak rotation velocity might be underestimated by the model, we indeed find that the difference in zero-point, $\Delta b$, with \citet{Miller11} shrinks by $\sim30$ per cent. 
If we now even turn to the purely observational `first order sTFR', this time using only the $z<1.3$ galaxies (122 SFGs) and the $v_{\rm rot, max}$ tracer, we find agreement to \citet{Miller11} ($\Delta b=0.02$). Again, we caution that this `first order' sample contains not only dispersion-dominated and merging galaxies, but also galaxies for which the maximum velocity is underestimated.
This exercise supports the interpretation that the disagreement with \cite{Miller11} is partly due to our selection of rotation-dominated systems. Beam-smearing corrections could lead to effects of comparable order, as is discussed in more detail in Appendix~\ref{selection-effects} and explicitly shown in Figure~\ref{fig:vcorr}.

The high$-z$ evolution of the bTFR has received less attention in the literature. 
At intermediate redshift ($z\sim1.2$), \cite{Vergani12} found no evolution of the bTFR when comparing to the local relation by \cite{McGaugh05}.
We compare our results to the slit-based relation at $z\sim2$ by \cite{Price16} using galaxies from the MOSDEF survey \citep{Kriek15}. 
\citet{Price16} use the $S_{0.5}=(0.5\cdot v_{\rm rot}^{2}+\sigma_{g}^{2})^{1/2}$ velocity tracer, which also incorporates dynamical support from disordered motions based on the assumption of isotropic (or constant) gas velocity dispersion $\sigma_{g}$ \citep{Weiner06a, Kassin07}.
\citet{Price16} show a plot of the $S_{0.5}-$bTFR of 178 SFGs, of which 35 (15) have detected (resolved) rotation measurements. For resolved galaxies, $S_{0.5}$ is obtained through combining a constant intrinsic velocity dispersion, and $v_{\rm rot,2.2}$. For unresolved galaxies, \citet{Price16} estimate $S_{0.5}$ through an rms velocity (see their Appendix~B for details). 
We use their fixed-slope fit ($a=1/0.39$) to compare their results to our 46 \kd galaxies at $z\sim2.3$ in the right panel of Figure~\ref{fig:stfr_lit}. Our fixed-slope fit is in agreement with the result by \citet{Price16} ($\Delta b=-0.03$). This is surprising at first, given the above discussion of IFS vs.\ slit-based rotation curve measurements, and the fact that the \citet{Price16}  sample contains a large fraction of objects without detected rotation. However, \citet{Price16} state that their findings regarding the S$_{0.5}$-bTFR do not change if they consider only the galaxies with detected rotation measurements. 
This is likely due to the detailed modelling and well-calibrated translation of line width to rotation velocity by the authors.
In general, any combination of velocity dispersion and velocity into a joined measure is expected to bring turbulent and even dispersion-dominated galaxies closer together in TFR space, which might further serve as an explanation for this good agreement \citep[see also][]{Covington10}.\footnote{
Partly, this is also the case for the measurements by \cite{Miller11, Miller12}, if a correction for turbulent pressure support is performed. Since their velocity dispersions are not available to us, however, only an approximative comparison is feasible. From this, we found agreement of their highest redshift bin ($z\sim1.5$) with our $0.6<z<2.6$ data in the $v_{\rm circ}$-sTFR plane, but still a significant offset at $z\sim1$.}

In summary, our inferred $v_{\rm rot}$-sTFR zero-points (i.e., not corrected for pressure support) agree with the work by \cite{Cresci09} and \cite{Tiley16}, but disagree with the work by \cite{Miller11}. Our $S_{0.5}$-bTFR zero-point agrees with the result by \cite{Price16}. 
We emphasize that the negligence of turbulent motions in the balance of forces leads to a relation which has lost its virtue to directly connect the baryonic kinematics to the central potential of the halo.\\

\section{TFR evolution in context}\label{discussion}

\subsection{Dynamical support of SFGs from $z\sim2.3$ to $z\sim0.9$}\label{evolv}

At fixed $v_{\rm circ}$, our sample shows higher $M_{\rm bar}$ and similar $M_{*}$ at $z\sim2.3$ as compared to $z\sim0.9$ (Figure~\ref{fig:stfr_btfr}). 
Galactic gas fractions are strongly increasing with redshift, as it has become clear in the last few years \citep{Tacconi10, Daddi10, Combes11, Genzel15, Tacconi17}. In our TFR sample, the baryonic mass of the $z\sim2.3$ galaxies is on average a factor of two larger as compared to $z\sim0.9$, while stellar masses are comparable.
The relative offset at fixed $v_{\rm circ}$ of our redshift subsamples in the bTFR plane, which is not visible in the sTFR plane, confirms the relevance of gas at high redshift.

Building on the recent work by \citetalias{WuytsS16} on the mass budgets of high$-z$ SFGs, we can identify through our Tully-Fisher analysis another redshift-dependent ingredient to the dynamical support of high$-z$ SFGs. The sTFR zero-point does not evolve significantly between $z\sim2.3$ and $z\sim0.9$. Since we know that there is less gas in the lower$-z$ SFGs, the `missing' baryonic contribution to the dynamical support of these galaxies as compared to $z\sim2.3$ has to be compensated by DM.
We therefore confirm with our study the increasing importance of DM to the dynamical support of SFGs (within $\sim1.3\,R_{e}$) through cosmic time. 
This might be partly due to the redshift dependence of the halo concentration parameters, which decrease with increasing redshift.
In the context of the toy model mentioned in Section~\ref{intro}, it is indeed the case that a decrease of the DM fraction as probed by the central galaxy with increasing redshift can flatten out or even reverse the naively expected, negative evolution of the TFR offset with increasing redshift. This will be discussed in more detail in Section~\ref{conclusions}.

The increase of baryon fractions with redshift is supported by other recent work: 
\citetalias{WuytsS16} find that the baryon fractions of SFGs within $R_{e}$ increase from $z\sim1$ to $z\gtrsim2$, with galaxies at higher redshift being clearly baryon-dominated \citep[see also][]{FS09, Alcorn16, Price16, Burkert16, Stott16, Contini16}. \citetalias{WuytsS16} also find that the baryonic mass fractions are correlated with the baryonic surface density within $R_{e}$, suggesting that the lower surface density systems at lower redshift are more diffuse and therefore probe further into the halo (consequently increasing their DM fraction).
Most recently, \cite{Genzel17} find in a detailed study based on the outer rotation curves of six massive SFGs at $z=0.9-2.4$ that the three $z>2$ galaxies are most strongly baryon-dominated.
On a statistical basis, this is confirmed through stacked rotation curves of more than 100 high$-z$ SFGs by \cite{Lang17}.

Given the average masses of our galaxies in the $YJ$ and $K$ subsamples, we emphasize that we are generally not tracing a progenitor-descendant population in our sample, since the average stellar and baryonic masses of the $z\sim2.3$ galaxies are already higher than for those at $z\sim0.9$ (Table~\ref{tab:properties}). 
It is very likely that a large fraction of the massive star-forming disk galaxies we observe at $z\gtrsim1$ have evolved into early-type galaxies (ETGs) by $z=0$, as discussed in the recent work by \cite{Genzel17}.
Locally, there is evidence that ETGs have high SFRs at early times, with the most massive ETGs forming most of their stars at $z\gtrsim2$ \citep[e.g.][]{Thomas10, McDermid15}. This view is supported by co-moving number density studies \citep[e.g.][]{Brammer11}, which also highlight that the mass growth of today's ETGs after their early and intense SF activity is mainly by the integration of (stellar) satellites into the outer galactic regions \citep{vDokkum10}. The observed low DM fractions of the massive, highest$-z$ SFGs seem to be consistent with the early assembly of local ETGs, with rapid incorporation of their baryon content.
In future work, we will compare our observations to semi-analytical models and cosmological zoom-in simulations to investigate in greater detail the possible evolutionary scenarios of our observed galaxies in the context of TFR evolution.

\subsection{Comparison to the local Universe}\label{comp_local}

In Figure~\ref{fig:stfr_btfr} we show the TFR zero-point evolution in context with the recent local studies by \cite{Reyes11} for the sTFR, and by \cite{Lelli16} for the bTFR. 
\citet{Reyes11} study the sTFR for a large sample of 189 disk galaxies, using resolved H$\alpha$ rotation curves. \citet{Lelli16} use resolved H{\textsc i} rotation curves and derive a bTFR for 118 disk galaxies.
To compare these local measurements to our high$-z$ \kd data, we assume that at $z\approx0$ the contribution from turbulent motions to the dynamical support of the galaxy is negligible, and therefore $v_{\rm circ}\equiv v_{\rm rot}$. We make the simplifying assumption that $v_{\rm circ}$ is comparable to $v_{80}$ and $v_{\rm flat}$ used by \citet{Reyes11} and \citet{Lelli16}, respectively (see \S~\ref{vel_measures} for a discussion). From \citet{Lelli16}, we use the fit to their subsample of 58 galaxies with the most accurate distances (see their classification).

For the sTFR as well as the bTFR we find significant offsets of the high$-z$ relations as compared to the local ones, namely $\Delta b_{\rm sTFR,z\sim0.9}=-0.44$, $\Delta b_{\rm sTFR,z\sim2.3}=-0.42$, $\Delta b_{\rm bTFR,z\sim0.9}=-0.44$ and $\Delta b_{\rm bTFR,z\sim2.3}=-0.27$. 
We have discussed in \S\S~\ref{tfr} and \ref{zp-ev} the zero-points of the `first order' TFRs as compared to our fiducial TFRs: while there is significant offset for both the `first order' sTFR and bTFR when comparing the $z\sim0.9$ and the $z\sim2.3$ subsamples, the overall offset to the local relations is reduced. The difference between the local relations and the full `first order' samples is only $\Delta b_{\rm sTFR}=-0.06$ and $\Delta b_{\rm bTFR}=0.02$, which would be consistent with no or only marginal evolution of the TFRs between $z=0$ and $0.6<z<2.3$.

For the interpretation of the offsets to the local relations, it is important to keep in mind that we measure the TFR evolution at the typical fixed circular velocity of galaxies in our high$-z$ sample. This traces the evolution of the TFR itself through cosmic time, not the evolution of individual galaxies. Our subsamples at $z\sim 0.9$ and $z\sim2.3$ are representative of the population of massive MS galaxies observed at those epochs, with the limitations as discussed in \S~\ref{tfr-sample}. Locally, however, the {\it typical} disk galaxy has lower circular velocity than our adopted reference velocity, and consequently lower mass (cf.\ e.g.\ Figure~1 by \citealp{Courteau15}). Figure~\ref{fig:stfr_btfr} does therefore not indicate how our galaxies will evolve on the TFR from $z\sim2$ to $z\sim0$, but rather shows how the relation itself evolves, as defined through the population of disk galaxies at the explored redshifts and mass ranges. 
This is also apparent if actual data points of low- and high-redshift disk galaxies are shown together. We show a corresponding plot for the bTFR in Appendix~\ref{alter-offset}.

In summary, our results suggest an evolution of the TFR with redshift, with zero-point offsets as compared to the local relations of $\Delta b_{\rm sTFR,z\sim0.9}=-0.44$, $\Delta b_{\rm sTFR,z\sim2.3}=-0.42$, $\Delta b_{\rm bTFR,z\sim0.9}=-0.44$ and $\Delta b_{\rm bTFR,z\sim2.3}=-0.27$. If galaxies with underestimated peak velocity, dispersion-dominated and disturbed galaxies are included, the overall evolution between the $z=0$ and $0.6<z<2.6$ samples is insignificant.

\subsection{The impact of uncertainties and model assumptions on the observed TFR evolution}\label{sum_uncertainties}

Before we interpret our observed TFR evolution in a cosmological context in Section~\ref{conclusions}, we discuss in the following uncertainties and modelling effects related to our data and methods. We find that uncertainties of mass estimates and velocities cannot explain the observed TFR evolution. Neglecting the impact of turbulent motions, however, could explain some of the tension with other work.

\subsubsection{Uncertainties of stellar and baryonic masses}\label{mass_uncertainties}

A number of approximations go into the determination of stellar and baryonic masses at high redshift. Simplifying assumptions like a uniform metallicity, a single IMF, or an exponentially declining SFH introduce significant uncertainties to the stellar age, stellar mass, and SFR estimates of high$-z$ galaxies. While the stellar mass estimates appear to be more robust against variations in the model assumptions, the SFRs, which are used for the molecular gas mass calculation, are affected more strongly \citep[see e.g.][for detailed discussions about uncertainties and their dependencies]{FS04, Shapley05, WuytsS07, WuytsS09, WuytsS16, Maraston10, Mancini11}. Most systematic uncertainties affecting stellar masses tend to lead to underestimates; if this were the case for our high$-z$ samples, the zero-point evolution with respect to local samples would be overestimated. However, the dynamical analysis by \citetalias{WuytsS16} suggests that this should only be a minor effect, given the already high baryonic mass fractions at high redshift.

An uncertainty in the assessment of gas masses at high redshift is the unknown contribution of atomic gas. In the local Universe, the gas mass of massive galaxies is dominated by atomic gas: for stellar masses of log$(M_{*}~[M_{\odot}])\approx10.5$, the ratio of atomic to molecular hydrogen is roughly $M_{\rm H{\textsc i}}/M_{\rm H_{2}}\sim3$ \citep[e.g.][]{Saintonge11}. While there are currently no direct galactic H{\textsc i} measurements available at high redshift,\footnote{
But see e.g.\ \cite{Wolfe05, Werk14} for measurements of H{\textsc i} column densities of the circum- and intergalactic medium using quasar absorption lines. From these techniques, a more or less constant {\it cosmological} mass density of neutral gas since at least $z\sim3$ is inferred \citep[e.g.][]{Peroux05, Noterdaeme09}. Recently, the need for a significant amount of non-molecular gas in the haloes of high$-z$ galaxies has also been invoked by the environmental study of the 3D-HST fields by \cite{Fossati17}.}
 a saturation threshold of the H{\textsc i} column density of only $\lesssim10\,M_{\odot}/{\rm pc}^{2}$ has been determined empirically for the local Universe \citep{Bigiel12}. The much higher gas surface densities of our high$-z$ SFGs therefore suggest a negligible contribution from atomic gas within $r\lesssim R_{e}$ \citepalias[see also][]{WuytsS16}. Consequently, the contribution of atomic gas to the maximum rotation velocity and to the mass budget within this radius should be negligible. However, there is evidence that locally H{\textsc i} disks are much more extended than optical disks \citep[e.g.][]{Broeils97}. If this is also true at high redshift, the total galactic H{\textsc i} mass fractions could still be significant at $z\sim1$, as is predicted by theoretical models \citep[e.g.][]{Lagos11, Fu12, Popping15}. Due to the lack of empirical confirmation, however, these models yet remain uncertain, especially given that they under-predict the observed high$-z$ molecular gas masses by factors of $2-5$. 
Within these limitations, we perform a correction for missing atomic gas mass at high$-z$ in our toy model discussion in Section~\ref{conclusions}.

Following \cite{Burkert16}, we have adopted uncertainties of $0.15$~dex for stellar masses, and $0.20$~dex for gas masses. This translates into an average uncertainty of $\sim0.15$~dex for baryonic masses.
These choices likely underestimate the systematic uncertainties in the error budget which can have a substantial impact on some of our results, because the slope as well as the scatter of the TFR are sensitive to the uncertainties.  For the presentation of our main results, we adopt local TFR slopes, thus mitigating these effects. 
In Appendix~\ref{uncertainties}, we explore the effect of varying mass uncertainties on free-slope fits of the TFR, together with implications on TFR residuals and evolution. We find that measurements of the zero-point are little affected by the uncertainties on mass, to an extent much smaller than the observed bTFR evolution between $z\sim2.3$ and $z\sim0.9$.

\subsubsection{Uncertainties of circular velocities}\label{vel_uncertainties}

We compute the uncertainties of the maximum circular velocity as the propagated errors on the {\it observed} velocity and $\sigma_{0}$, including an uncertainty on $q$ of $\sim20$ per cent. The latter is a conservative choice in the light of the current \kd magnitude cut of $Ks<23$ \citep[cf.][]{vdWel12}. 
For details about the observed quantities, see \citetalias{Wisnioski15}, and \citetalias{WuytsS16} for a comparison between observed and modelled velocities and velocity dispersions. 
The resulting median of the propagated circular velocity uncertainty is 20~km/s.

Maximum circular velocities can be systematically underestimated: although the effective radius enters the modelling procedure as an independent constraint, the correction for pressure support can lead to an underestimated turn-over radius if the true turn-over radius is not covered by observations. 
For our TFR sample we selected only galaxies where modelled and observed velocity and dispersion profiles are in good agreement, and where the maximum or flattening of the rotation curve is covered by observations. It is therefore unlikely that our results based on the TFR sample are affected by systematic uncertainties of the maximum circular velocity.

\subsubsection{Effects related to different velocity measures and models}\label{vel_measures}

The different rotation velocity models and measures used in the literature might affect comparisons between different studies. 
Some TFR studies adopt the rotation velocity at 2.2 times $R_{d}$, $v_{2.2}$, as their fiducial velocity to measure the TFR. We verified that for the dynamical modelling as described above, $v_{\rm{circ, 2.2}}$ equals $v_{\rm{circ, max}}$, and $v_{\rm{rot, 2.2}}$ equals $v_{\rm{rot, max}}$ with an average accuracy of $\lesssim1$~km/s. 
Other commonly used velocity measures are $v_{\rm flat}$, $v_{3.2}$, and $v_{80}$, the rotation velocity at the radius which contains 80 per cent of the stellar light. For a pure exponential disk, this corresponds to roughly $v_{3.0}$ \citep{Reyes11}. It has been shown by \cite{Hammer07} that $v_{\rm flat}$ and $v_{80}$ are comparable in local galaxies. For the exponential disk model including pressure support which we use in our analysis, $v_{\rm{rot(circ), max}}$ is on average $\lesssim15(10)$~km/s larger than $v_{\rm{rot(circ), 3.2}}$. Since $v_{3.2}$ and $v_{80}$ are, however, usually measured from an `arctan model' with an asymptotic maximum velocity \citep{Courteau97}, reported values in the literature generally do {\it not} correspond to the respective values at these radii from the thick exponential disk model with pressure support. \cite{Miller11} show that for their sample of SFGs at $0.2<z<1.3$, the typical difference between $v_{2.2}$ and $v_{3.2}$, as computed from the arctan model, is on the order of a few per cent \citep[see also][]{Reyes11}. 
This can also be assessed from Figure~6 by \cite{Epinat10}, who show examples of velocity fields and rotation curves for different disk models (exponential disk, isothermal sphere, `flat', arctan). By construction, the peak velocity of the exponential disk is higher than the arctan model rotation velocity at the corresponding radius. 

We conclude that our TFR `velocity' values derived from the peak rotation velocity of a thick exponential disk model are comparable to $v_{\rm flat}$, and close to $v_{3.2}$ and $v_{80}$ from an arctan model, with the limitations outlined above. The possible systematic differences of $<20$~km/s between the various velocity models and measures cannot explain the observed evolution between $z=0$ and $0.6<z<2.6$.

Another effect on the shape of the velocity and velocity dispersion profiles is expected if contributions by central bulges are taken into account. We have tested for a sample of more than 70 galaxies that the effect of including a bulge on our adopted velocity tracer, $v_{\rm circ, max}$ is on average no larger than 5 per cent. From our tests, we do not expect the qualitative results regarding the TFR evolution between $z\sim2.3$ and $z\sim0.9$ presented in this paper to change if we include bulges into the modelling of the mass distribution.

\subsubsection{The impact of turbulent motions}\label{turbulent}

The dynamical support of star-forming disk galaxies can be quantified through the relative contributions from ordered rotation and turbulent motions \citep[see also e.g.][]{Tiley16}. We consider only rotation-dominated systems in our TFR analysis, namely galaxies with $v_{\rm rot, max}/\sigma_{0}>\sqrt{4.4}$. Because of this selection, the effect of $\sigma_{0}$ on the velocity measure is already limited, with median values of $v_{\rm rot,max}=233$~km/s at $z\sim0.9$, and $239$~km/s at $z\sim2.3$, {\it vs.\ }median values of $v_{\rm circ,max}=239$ and $v_{\rm circ,max}=260$~km/s at $z\sim0.9$ and $z\sim 2.3$, respectively (Table~\ref{tab:properties}).

However, this difference translates into changes regarding e.g.\ the TFR scatter: for the $v_{\rm rot,max}$-TFR, we find a scatter of $\zeta_{\rm int, sTFR}=0.28$ and $\zeta_{\rm int, bTFR}=0.31$ at $z\sim0.9$, and at $z\sim2.3$ we find $\zeta_{\rm int, sTFR}=0.33$ and $\zeta_{\rm int, bTFR}=0.33$, with those values being consistently higher than the values reported for the $v_{\rm circ,max}$-TFR sample in Table~\ref{tab:fits}. 
More significantly, neglecting the contributions from turbulent motions affects the zero-point evolution: without correcting $v_{\rm rot, max}$ for the effect of pressure support, we would find $\Delta b_{\rm sTFR,z\sim0.9}=-0.34$, $\Delta b_{\rm sTFR,z\sim2.3}=-0.26$, $\Delta b_{\rm bTFR,z\sim0.9}=-0.33$ and $\Delta b_{\rm bTFR,z\sim2.3}=-0.09$. The inferred zero-points at higher redshift are affected more strongly by the necessary correction for pressure support (cf.\ Figure~\ref{fig:stfr_btfr}).

These results emphasize the increasing role of pressure support with increasing redshift, confirming previous findings by e.g.\ \cite{FS09, Epinat09, Kassin12}; \citetalias{Wisnioski15}.
It is therefore clear that turbulent motions must not be neglected in kinematic analyses of high$-z$ galaxies. If the contribution from pressure support to the galaxy dynamics is dismissed, this will lead to misleading conclusions about TFR evolution in the context of high$-z$ and local measurements.\\

\section{A toy model interpretation}\label{conclusions}

The relative comparison of our $z\sim2.3$ and $z\sim0.9$ data and local relations indicates a non-monotonic evolution of the bTFR zero-point with cosmic time (Figure~\ref{fig:stfr_btfr}). In this section, we present a toy model interpretation of our results, aiming to explain the redshift evolution of both the sTFR and the bTFR, in particular the relative zero-point offsets at $z\sim2.3$, $z\sim0.9$, and $z\sim0$.
  
The basic premise is that galaxies form at the centers of DM haloes. A simple model for a DM halo in approximate equilibrium is a truncated isothermal sphere, limited by the radius $R_{h}$ where the mean density equals 200 times the critical density of the Universe. The corresponding redshift-dependent relations between halo radius, mass $M_{h}$, and circular velocity $V_{h}$ are
\begin{equation}\label{eq:virial}
	M_{h}=\frac{V_{h}^{3}}{10G\cdot H(z)} \hspace{5mm} {\rm ;} \hspace{5mm} R_{h}=\frac{V_{h}}{10H(z)}
\end{equation}
\citep{MMW98}, where $H(z)$ is the Hubble parameter, and $G$ is the gravitational constant. The first equation shows that the relation between $M_{h}$ and $V_{h}$ is a smooth function of redshift.

In theory, the relation between these halo properties and corresponding galactic properties can be complex due to the response of the halo to the formation of the central galaxy \citep[see e.g.\ the discussions on halo contraction {\it vs.\ }expansion by][]{Duffy10, Dutton16, Velliscig14}. 
However, recent studies and modelling of high$-z$ SFGs now provide a number of empirical constraints that implicitly contain information on the DM halo profile on galactic scales.

Relations corresponding to Equations~\eqref{eq:virial} for the central baryonic galaxy can then be derived by assuming a direct mapping between the halo and galaxy mass and radius. Information on the inner halo profile is contained in parameters such as the disk mass fraction $m_{d}=M_{\rm bar}/M_{h}$, or the central DM fraction $f_{\rm DM}(r)=v^2_{\rm DM}(r)/v^2_{\rm circ}(r)$. 
For our galaxies, we know their stellar mass $M_{*}$ and effective radius $R_{e}$, their baryonic mass $M_{\rm bar}$ and gas mass fraction $f_{\rm gas}=M_{\rm gas}/M_{\rm bar}$ from empirical scaling relations, and their circular velocity $v_{\rm circ}(r)$ and related central DM fraction $f_{\rm DM}(r)$ from dynamical modelling, as detailed in \S\S~\ref{mass} and \ref{modelling} and in the references given there. 
We further have an estimate of their average baryonic disk mass fraction $m_{d}$ \citep{Burkert16}. 
We can combine this information to construct a toy model of the TFR zero-point evolution, where we take the redshift dependencies of these various parameters into account (see Appendix~\ref{model-theory} for a detailed derivation):
\begin{equation}\label{eq:btfr}
	M_{\rm bar}=\frac{v_{\rm circ}^{3}(R_{e})}{H(z)} \cdot \frac{[1-f_{\rm DM}(R_{e},z)]^{3/2}}{m_{d}^{1/2}(z)}\cdot C
\end{equation}
\begin{equation}\label{eq:stfr}
	M_{*}=\frac{v_{\rm circ}^{3}(R_{e})}{H(z)} \cdot \frac{[1-f_{\rm DM}(R_{e},z)]^{3/2}\,[1-f_{\rm gas}(z)]}{m_{d}^{1/2}(z)}\cdot C^{\prime},
\end{equation}
where $C$ and $C^{\prime}$ are constants. 
Here, we have assumed that, in contrast to the disk mass fraction, the proportionality factor between DM halo radius and galactic radius is independent of redshift \citep[see e.g.][]{Burkert16}.

Equations~\eqref{eq:btfr} and \eqref{eq:stfr} reveal that the TFR evolution can be strongly affected by changes of $f_{\rm DM}(R_{e})$, $m_{d}$, or $f_{\rm gas}$ with redshift, and does not necessarily follow the smooth evolution of the halo parameters given in Equation~\eqref{eq:virial}. 
There have been indications for deviations from a simple smooth TFR evolution scenario in the theoretical work by \cite{Somerville08}. Also the recent observational compilation by \cite{Swinbank12} showed a deviating evolution (although qualified as consistent with the smooth evolution scenario).

Evaluating Equations~\eqref{eq:btfr} and \eqref{eq:stfr} at fixed $v_{\rm circ}(R_{e})$, we learn the following: 
(i) if $f_{\rm DM}(R_{e})$ decreases with increasing redshift, the baryonic and stellar mass will increase and consequently the TFR zero-point will increase;
(ii) if $m_{d}$ increases with increasing redshift, the baryonic and stellar mass will decrease and consequently the TFR zero-point will decrease;
(iii) if $f_{\rm gas}$ increases with increasing redshift, the stellar mass will decrease and consequently the sTFR zero-point will decrease.
These effects are illustrated individually in Figure~\ref{fig:tfrev_effects} in Appendix~\ref{model-derivation}. 

\begin{figure*}
\centering
	\includegraphics[width=0.9\textwidth]{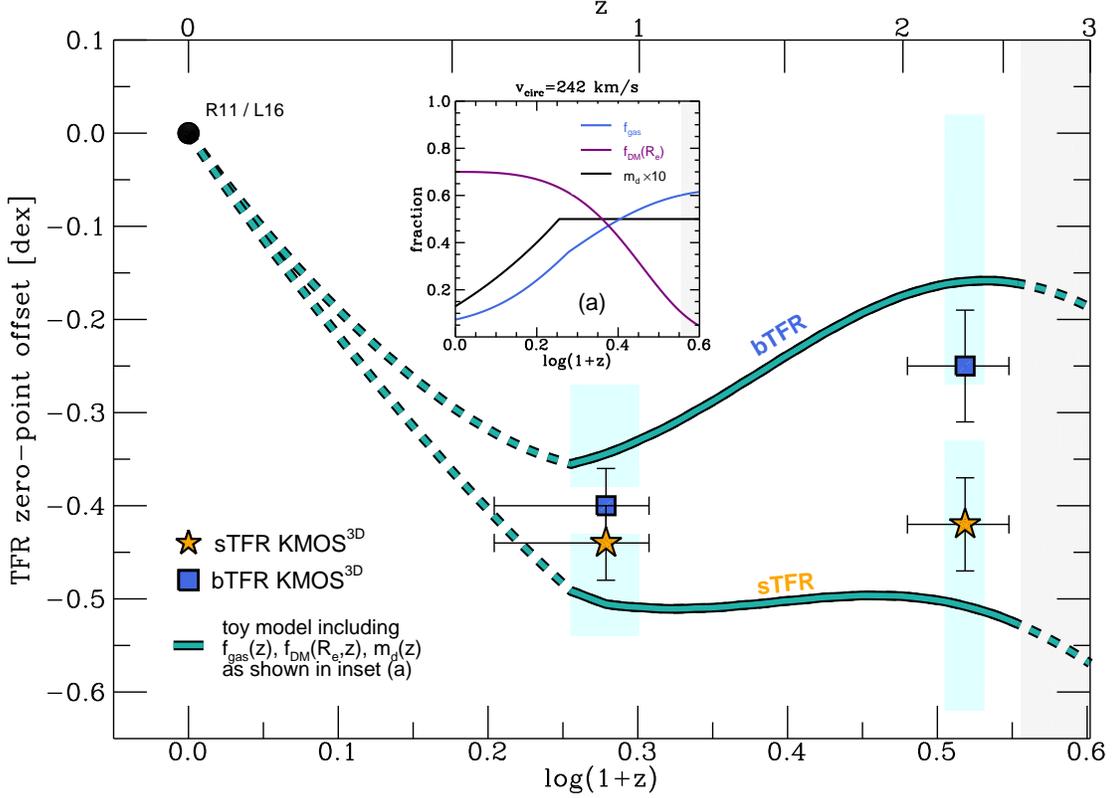}
    \caption{TFR zero-point offsets of the stellar and baryonic mass TFRs as a function of cosmic time. The \kd data is shown as yellow stars (sTFR) and blue squares (bTFR), in relation to the corresponding local normalizations by \citeauthor{Reyes11}\ (2011; R11) and \citeauthor{Lelli16}\ (2016b; L16). 
    The horizontal error bars of the \kd data points indicate the spanned range in redshift, while the vertical error bars show fit uncertainties. 
    The bTFR data points are corrected for neglected atomic gas at $z\sim0.9$ and $z\sim2.3$, as detailed in the main text.
    The green dashed and solid lines show predictions for the bTFR and sTFR evolution from our toy model (Equations~\eqref{eq:btfr} and \eqref{eq:stfr}). This model takes into account the empirically motivated redshift dependencies of $f_{\rm gas}$, $f_{\rm DM}(R_{e})$, and $m_{d}$, in particular as they are shown in inset (a).  Regions in redshift space where the model is not well constrained due to a lack of observational constraints in particular on $m_{d}$ are indicated as dashed lines. Observational constraints come from \cite{Saintonge11} and \cite{Tacconi17} for $f_{\rm gas}(z)$, from \cite{Martinsson13a, Martinsson13b} and \citetalias{WuytsS16} for $f_{\rm DM}(R_{e},z)$, and from \cite{Moster13} and \cite{Burkert16} for $m_{d}(z)$, as detailed in Appendix~\ref{model-observations}.
    Our proposed parametrizations are valid only up to $z\approx 2.6$, as indicated by the grey shading in the main figure and inset (a). 
    As cyan shaded areas we indicate by way of example how the model TFR evolution would change if DM fractions would be higher/lower by 0.1 at $z=0$, $z=0.9$, and $z=2.3$ (horizontal ranges are $\pm0.1 z$).
    The observed TFR evolution is reasonably matched by a model where the disk scale length is proportional to the halo radius, and where $f_{\rm gas}$ and $m_{d}$ increase with redshift, while $f_{\rm DM}(R_{e})$ decreases with redshift.}
    \label{fig:context}
\end{figure*}
 
We constrain our toy model at redshifts $z=0$, $z\sim0.9$, and $z\sim2.3$ as follows: 
the redshift evolution of $f_{\rm gas}$ is obtained through the empirical atomic and molecular gas mass scaling relations by \cite{Saintonge11} and \cite{Tacconi17}. At fixed circular velocity, $f_{\rm gas}$ evolves significantly with redshift, where $z\sim2$ galaxies have gas fractions which are about a factor of eight higher than in the local Universe. 
The redshift evolution of $f_{\rm DM}(R_{e})$ is constrained through the observational results by \cite{Martinsson13a, Martinsson13b} in the local Universe, and by \citetalias{WuytsS16} at $z\sim0.9$ and $z\sim2.3$. We tune the redshift evolution of $f_{\rm DM}(R_{e})$ within the ranges allowed by these observations to optimize the match between the toy model and the observed TFR evolution presented in this paper. $f_{\rm DM}(R_{e})$ evolves significantly with redshift, with $z\sim2$ DM fractions which are about a factor of five lower than at $z=0$.
$m_{d}$ is constrained by the abundance matching results by \cite{Moster13} in the local Universe, whereas at $0.8<z<2.6$ we adopt the value deduced by \cite{Burkert16}. 
Details on the parametrization of the above parameters are given in Appendix~\ref{model-observations}.

In Figure~\ref{fig:context} we show how these empirically motivated, redshift-dependent DM fractions, disk mass fractions, and gas fractions interplay in our toy model framework to approximately explain our observed TFR evolution, specifically the TFR zero-point offsets at fixed circular velocity as a function of cosmic time. In particular, this is valid at $z=0$, $z=0.9$, and $z=2.3$, while we have partially interpolated in between. 
Our observed \kd TFR zero-points of the bTFR (blue squares) and the sTFR (yellow stars) at $z\sim0.9$ and $z\sim2.3$ are shown in relation to the local TFRs by \cite{Lelli16} and \cite{Reyes11}. 
The horizontal error bars of the \kd data points indicate the spanned range in redshift, while the vertical error bars show fit uncertainties. 
For this plot, we also perform a correction for atomic gas at high redshift:\footnote{\cite{Lelli16} neglect molecular gas for their bTFR, but state that it has generally a minor dynamical contribution.} 
we follow the theoretical prediction that, at fixed $M_{*}$, the ratio of atomic gas mass to stellar mass does not change significantly with redshift \citep[e.g.][]{Fu12}. We use the fitting functions by \cite{Saintonge11} to determine the atomic gas mass for galaxies with log$(M_{*}~[M_{\odot}])=10.50$, which corresponds to the average stellar mass of our TFR galaxies at $v_{\rm ref}=242$~km/s in both redshift bins. We find an increase of the zero-point of $+0.04$~dex at $z\sim0.9$ and $+0.02$~dex at $z\sim2.3$. This is included in the figure.

We show as green lines our empirically constrained toy model governed by Equations~\eqref{eq:btfr} and \eqref{eq:stfr}. This model assumes a redshift evolution of $f_{\rm gas}$, $f_{\rm DM}(R_{e})$, and $m_{d}$ as shown by the blue, purple, and black lines, respectively, in inset (a) in Figure~\ref{fig:context} (details are given in Appendix~\ref{model-observations}). 
In this model, the increase in $f_{\rm gas}$ is responsible for the deviating (and stronger) evolution of the sTFR as compared to the bTFR. 
The decrease of $f_{\rm DM}(R_{e})$ is responsible for the upturn/flattening of the bTFR/sTFR evolution. 
The increase of $m_{d}$ leads to a TFR evolution which is steeper than what would be expected from a model governed only by $H(z)$ (see also Fig.~\ref{fig:tfrev_effects}). 
Our toy model evolution is particularly sensitive to changes of $f_{\rm DM}(R_{e})$ with redshift. We illustrate this by showing as cyan shaded areas in Figure~\ref{fig:context} how the toy model evolution would vary if we would change only $f_{\rm DM}(R_{e})$ by $\pm0.1$ at $z=0$, $z=0.9$, and $z=2.3$.

We note that the toy model zero-point offset at $R_{e}$ as derived from Equations~\eqref{eq:btfr} and \eqref{eq:stfr}, and based on a thin exponential baryon distribution, is comparable to our empirical TFR offset for a thick exponential disk and using $v_{\rm circ, max}$, since the correction factors for the circular velocity measure from thin to thick exponential disk, and from $v_{\rm circ}(R_{e})$ to $v_{\rm circ,max}\approx v_{\rm circ}(r_{2.2})$, are both of the order of $\sim5$ per cent and approximately compensate one another. The toy model slope ($a=3$) is shallower than our adopted local slopes. In Appendix~\ref{uncertainties} we show that the usage of a reference velocity leads to negligible zero-point differences of TFR fits with different slopes.

Although our toy model is not a perfect match to the observed TFR evolution, it reproduces the observed trends reasonably well: for the sTFR, the zero-point decreases from $z=0$ to $z\sim1$, but there is no or only marginal evolution between $z\sim1$ and $z\sim2$. In contrast, there is a significantly non-monotonic evolution of the bTFR zero-point, such that the zero-point first decreases from $z=0$ to $z\sim1$, and then increases again up to $z\sim2$. 
We note that although we show the TFR evolution up to $z=3$, the constraints on $f_{\rm DM}(R_{e})$ and $m_{d}$ are valid only up to $z\approx2.6$, as indicated in the figure by the grey shading. Also in the redshift range $0\lesssim z\lesssim 0.8$ the model is poorly constrained because we assume a simplistic evolution of $m_{d}$ (cf.\ Appendix~\ref{model-observations}).

A more complete interpretation of our findings also at intermediate redshift has to await further progress in observational work. 
With the extension of the \kd survey towards lower mass galaxies and towards a more complete redshift coverage in the upcoming observing periods, we might already be able to add in precision and redshift range to our model interpretation.
Our current data and models, however, already show the potential of state-of-the-art high$-z$ studies of galaxies to constrain parameters which are important also for theoretical work.

We would like to caution that our proposed model certainly draws a simplified picture.
For instance, the assumption of a common scale length of the atomic gas as well as the molecular gas plus stars, as we did for this exercise, can only be taken as approximate, given the high central surface mass densities of our typical high$-z$ galaxies (see \S~\ref{mass_uncertainties}, and \citetalias{WuytsS16}). 
Also, the effective radii predicted by our ``best fit'' toy model are 10-30 per cent larger than what is observed. 
Other factors not addressed in our approach might also come into play: we did not explore in detail the possible effects of varying halo spin parameter $\lambda$ or of the ratio between baryonic and DM specific angular momenta $j_{\rm bar}/j_{\rm DM}$, which commonly relate $R_{h}$ to $R_{d}$. 
We also note that possible conclusions on the NFW halo concentration parameter $c$ are in tension with current models (cf.\ Appendix~\ref{model-observations}). 
We therefore caution that our proposed toy model perspective can only reflect general trends, in particular the relative TFR zero-point offsets at $z=0$, $z=0.9$, and $z=2.3$, and likely misses other relevant ingredients.

Having in mind the limitations outlined above, we conclude that the observed evolution of the mass-based TFRs can be explained in the framework of virialized haloes in an expanding $\Lambda$CDM universe, with galactic DM fractions, disk mass fractions, and gas fractions that are evolving with cosmic time. 
Adopting the proposed evolution of the model parameters in Equations~\eqref{eq:btfr} and \eqref{eq:stfr} as described above and shown in inset (a) in Figure~\ref{fig:context}, namely at fixed $v_{\rm circ}$ increasing $f_{\rm gas}$ and $m_{d}$, and decreasing $f_{\rm DM}(R_{e})$ with redshift, leads to a redshift evolution of the TFR which is non-monotonic, in particular for the bTFR.\\

\section{Summary}\label{summary}

We have investigated the mass-based Tully-Fisher relations (TFRs) of massive star-forming disk galaxies between redshift $z\sim2.3$ and $z\sim0.9$ as part of the \kd survey. 
All our data are reduced and analyzed in a self-consistent way. The spatially resolved nature of our observations enables reliable modelling of individual galaxies, and allows for a careful selection of objects based on kinematic properties and data quality. We have taken into account inclination, beam-smearing, and instrumental broadening, and we have incorporated the significant effects of pressure support to the gravitational potential at these redshifts in our derivation of the circular velocities.

We find that the TFR is clearly in place already at $0.6<z<2.6$ (\S~\ref{tfr}). Its scatter increases with redshift, but we did not find any second-order parameter dependencies when adopting a local slope. At fixed $v_{\rm circ,max}$, we find higher $M_{\rm bar}$ but similar $M_{*}$ at $z\sim2.3$ as compared to $z\sim0.9$ (\S~\ref{zp-ev}). 
This highlights the important effects of the evolution of $f_{\rm gas}$, where, at the same stellar mass, high$-z$ star-forming galaxies (SFGs) have significantly higher gas fractions than lower$-z$ SFGs. This strengthens earlier conclusions by \cite{Cresci09} in the context of the interpretation of TFR evolution.
Since we do not find a significant evolution of the sTFR between $z\sim2.3$ and $z\sim0.9$, our observed TFR evolution together with the decrease of $f_{\rm gas}$ with decreasing redshift, implies that the contribution of dark matter (DM) to the dynamical mass on the galaxy scale has to increase with decreasing redshift to maintain the dynamical support of the galaxy as measured through $v_{\rm circ,max}$. 
Our results complement the findings in other recent work that higher$-z$ SFGs are more baryon-dominated (\S~\ref{evolv}).

Comparing to other selected high$-z$ TFR studies, we find agreement with the work by \cite{Cresci09, Price16, Tiley16}, but disagreement with the work by \cite{Miller11} (\S~\ref{comp_high}).
The significant differences in zero-point offsets of our high$-z$ TFRs as compared to the local relations by \cite{Reyes11} and \cite{Lelli16} indicate an evolution of the TFR with cosmic time (\S~\ref{comp_local}). From the local Universe to $z\sim0.9$ and further to $z\sim2.3$, we find a non-monotonic TFR zero-point evolution which is particularly {\rm pronounced} for the bTFR.

To explain our observed TFR evolution, we present a toy model interpretation guided by an analytic model of disk galaxy evolution (Section~\ref{conclusions}). This model takes into account empirically motivated gas fractions, disk mass fractions, and central DM fractions with redshift. We find that the increasing gas fractions with redshift are responsible for the increasingly deviating evolution between the sTFR and the bTFR with redshift. The decreasing central DM fractions with redshift result in the flattening/upturn of the sTFR/bTFR zero-point evolution at $0.9<z<2.3$.
This simple model matches our observed TFR evolution reasonably well. 

It will be interesting to make more detailed comparisons between the growing amount of observations that can constrain the TFR at high redshift, and the newest generation of simulations and semi-analytical models.
Further investigations of galaxies at lower ($z\lesssim0.7$) and higher ($z\gtrsim2.5$) redshifts using consistent reduction and analysis techniques will help to unveil the detailed evolution of the mass-based TFR, and to reconcile current tensions in observational work. Another important quest is to provide data which cover wider ranges in velocity and mass at these high redshifts to minimize uncertainties in the fitting of the data, and to investigate if the TFR slope changes with redshift.\\


\acknowledgements 
We are grateful to the anonymous referee for a constructive report which helped to improve the manuscript.
We thank the ESO Paranal staff for their helpful support with the KMOS observations for this work.
We are grateful to Jonathan Freundlich, Susan Kassin, Federico Lelli, Raymond Simons, Jakob Walcher, and in particular to Amiel Sternberg and Simon White, for fruitful discussions, and to Rachel Somerville, Mike Williams, and Dennis Zaritsky for valuable insight into various aspects of this work.
We thank Sedona Price for providing us with details on the fits by \cite{Price16}, and we are grateful to Brandon Kelly and Marianne Vestergaard for providing us with a modified version of the Bayesian approach to linear regression code \citep{Kelly07} which allows for fixing the slope. 
JC acknowledges the support of the Deutsche Zentrum f\"ur Luft- und Raumfahrt (DLR) via Project ID 50OR1513.
MF and DJW acknowledge the support of the Deutsche Forschungsgemeinschaft (DFG) via Projects WI~3871/1-1, and WI~3871/1-2.\\



\appendix

\section{The effects of sample selection}\label{selection-effects}

\begin{figure}
\centering
 	\includegraphics[width=\columnwidth]{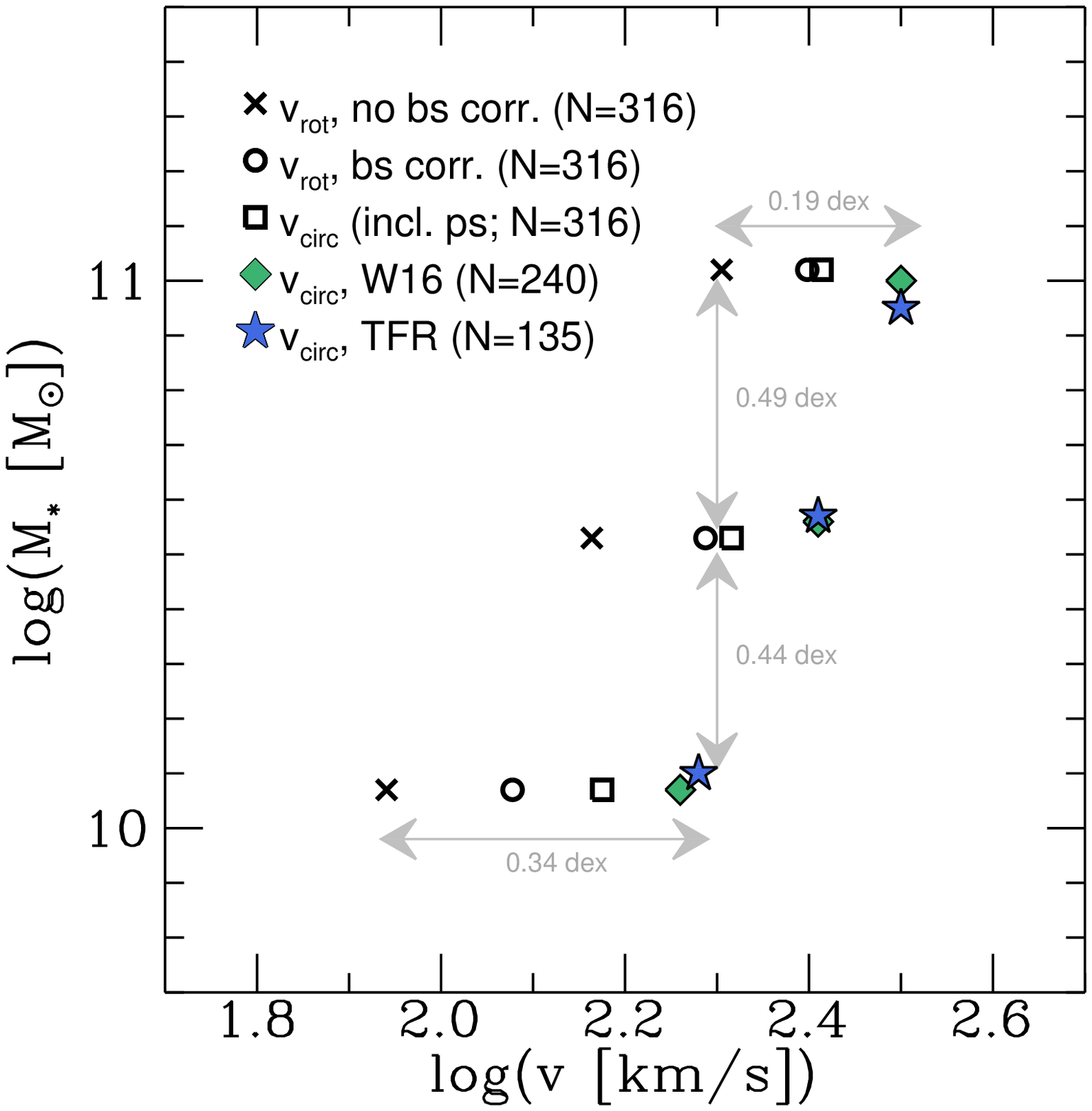}
   \caption{Illustration of different correction (black symbols) and selection (colored symbols) effects on the mean maximum rotation, or circular, velocity for three stellar mass bins, log$(M_{*}~[M_{\odot}])$<10.3, 10.3<log$(M_{*}~[M_{\odot}])$<10.8, and 10.8<log$(M_{*}~[M_{\odot}])$. Black crosses show the observed maximum velocity corrected for inclination but not beam-smearing. Black circles include the beam-smearing correction. Black squares include the correction for pressure support, leading to the maximum circular velocity as defined in Equation~\eqref{eq:vcirc}. These data points consider all resolved \kd galaxies. The corresponding mean circular velocities for the \citetalias{WuytsS16} sample are shown as green diamonds, and the final TFR sample is shown as blue stars. 
   The final selection steps for our TFR sample detailed in \S~\ref{tfr-sample} have a much smaller effect than the beam-smearing and pressure support correction, and than the selection of galaxies suited for a kinematic disk modelling.}
    \label{fig:vcorr}
\end{figure}

\begin{figure*}
	\centering
	\includegraphics[width=\columnwidth]{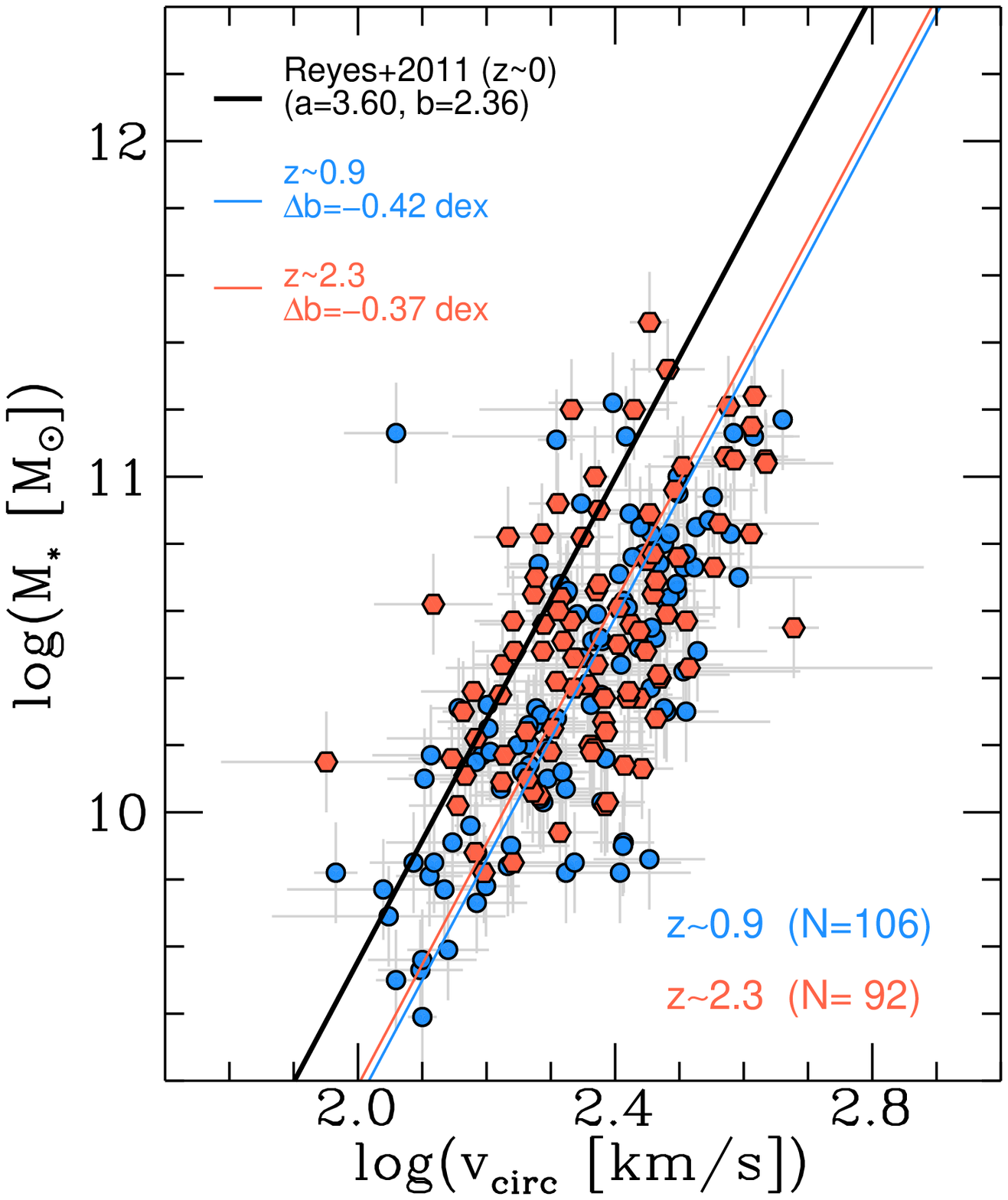}
	\hspace{8mm}
	\includegraphics[width=\columnwidth]{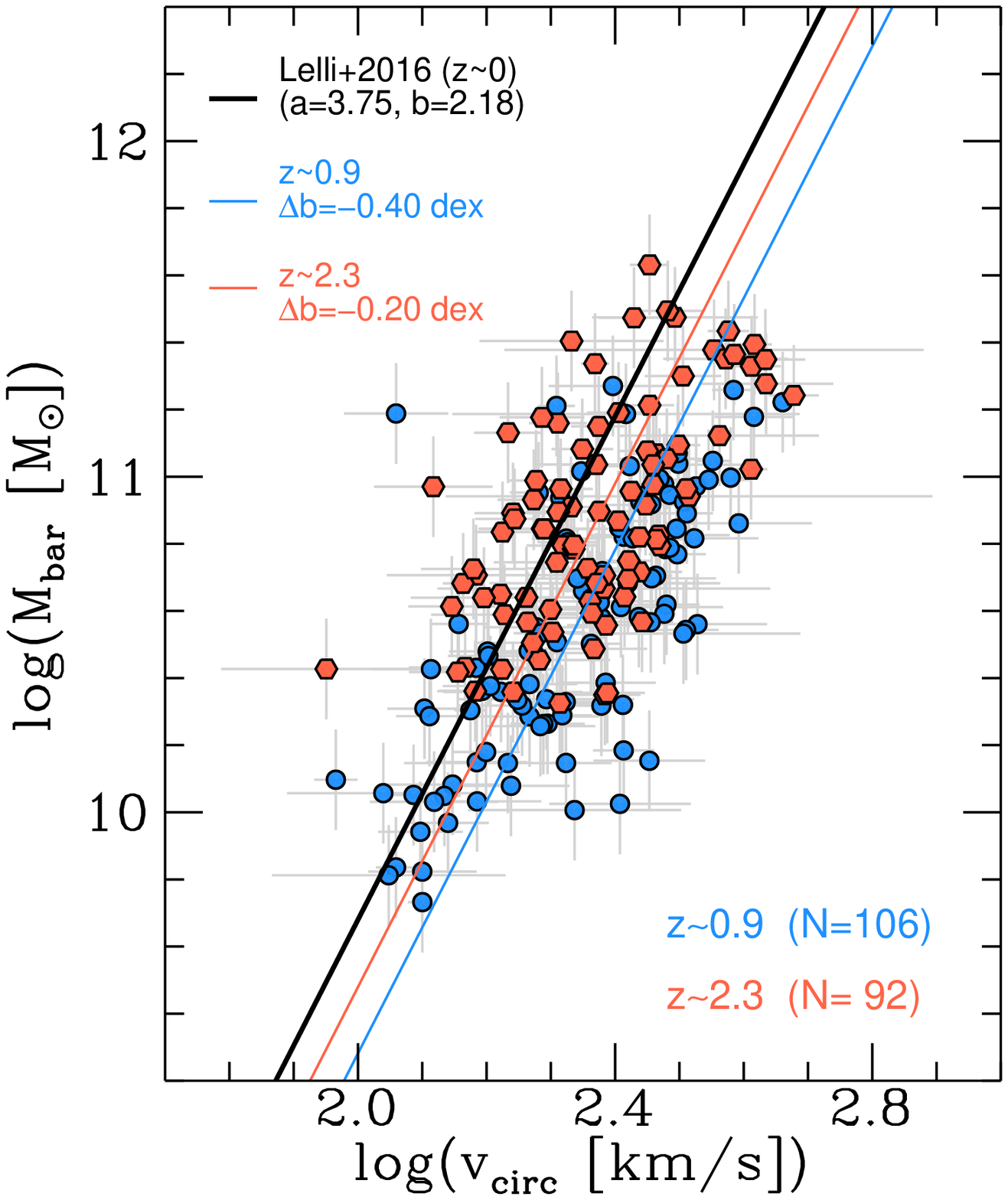}
   \caption{Fixed-slope fits for the sTFR (left) and the bTFR (right) using local (black) slopes to the \citetalias{WuytsS16} subsamples at $z\sim0.9$ (blue) and $z\sim2.3$ (red).
   We find no (or only marginal) evolution of the sTFR zero-point in the studied redshift range, but significant evolution of the bTFR given the typical fit uncertainties of $\delta b=0.05$~dex. While there are changes of up to $+0.07$~dex when comparing to the TFR sample evolution (Figure~\ref{fig:stfr_btfr}), mostly due to underestimated velocities when the maximum of the rotation curve is not covered by data, we see the same general trends as for the refined TFR sample.}
    \label{fig:w16_tfr}
\end{figure*}

For the discussion of the TFR at high redshift it is important to be aware not only of the location of the subsample of `TFR galaxies' within a larger parent sample, but also of the effect of the necessary corrections to the observed velocity which ultimately lead to the high-$z$ TFR.
Figure~\ref{fig:vcorr} illustrates for three stellar mass bins (log$(M_{*}~[M_{\odot}])$<10.3; 10.3<log$(M_{*}~[M_{\odot}])$<10.8; 10.8<log$(M_{*}~[M_{\odot}])$) how the mean maximum rotation velocity changes through corrections for beam-smearing and pressure support, when selecting for rotating disks, and when eventually selecting for `TFR galaxies' following the steps outlined in \S~\ref{tfr-sample}. 

The effect of beam-smearing on the rotation velocity is with differences of $\gtrsim0.1$~dex significant for our galaxies, translating into an offset in stellar mass of $\gtrsim0.4$~dex. Considering next the impact of turbulent motions, one can clearly see how this is larger for lower-mass (and lower-velocity) galaxies.\footnote{Taking turbulent motions into account also has a larger effect at higher redshift due to the increase of intrinsic velocity dispersion with redshift. This is not explicitly shown in Figure~\ref{fig:vcorr}.} 
This reflects the larger proportion of dispersion-dominated systems at masses of log$(M_{*}~[M_{\odot}])\lesssim10$. Correcting the observed rotation velocity for these two effects does not involve a reduction of the galaxy sample, and the corresponding data points in Figure~\ref{fig:vcorr} include all 316 resolved \kd galaxies.
The procedure of selecting galaxies suitable for a kinematic disk modelling (\citetalias{WuytsS16}; \S~\ref{tfr-sample}) has a noticeable effect in the full mass range explored here. 
It becomes clear that the further, careful selection of galaxies best eligible for a Tully-Fisher study has an appreciable effect on the mean velocity of about $0.02-0.03$~dex, but is minor as compared to the other effects discussed.

While we consider the selection of the `TFR sample' important due to the $v_{\rm rot,max}/\sigma_{0}$ cut and the reliable recovery of the true maximum rotation velocity, we note that it only leads to a small change in TFR parameters as compared to the \citetalias{WuytsS16} sample (Figure~\ref{fig:w16_tfr}).\\

\section{An alternative method to investigate TFR evolution}\label{alter-offset}

It is standard procedure in investigations of the TFR to adopt a local slope for galaxy subsamples in different redshift bins, and to quantify its evolution in terms of zero-point variations, since high$-z$ samples often span too limited a range in mass and velocity to reliably constrain a slope. 
This method has two shortcomings: first, potential changes in slope with cosmic time are not taken into account. Second, every investigation of TFR evolution is tied to the adopted slope which sometimes complicates comparative studies.

\begin{figure}
	\centering
	\includegraphics[width=\columnwidth]{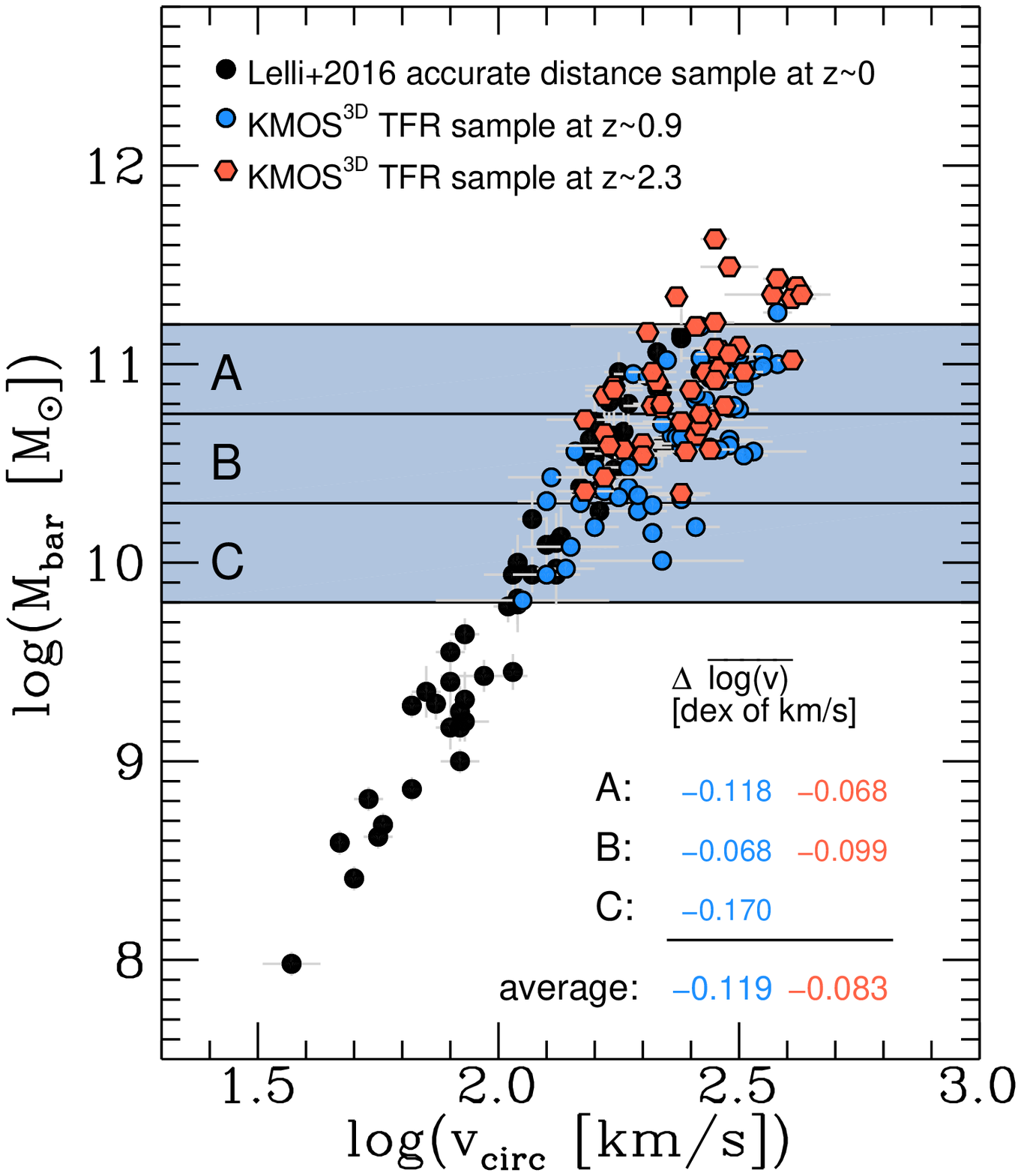}
    \caption{Our TFR galaxies at $z\sim2.3$ (red) and $z\sim0.9$ (blue) together with the local sample by \cite{Lelli16} (black) in the bTFR plane. We calculate weighted mean velocities of the redshift subsamples in the three mass bins labelled `A', `B', and `C', in order to investigate the TFR evolution in a way independent of the usual functional form of the TFR. The velocity differences averaged over the mass bins of $\Delta \overline{{\rm log}(v_{\rm circ}~[{\rm km/s}])}=-0.119$ between $z\sim0.9$ and $z=0$, and of $\Delta \overline{{\rm log}(v_{\rm circ}~[{\rm km/s}])}=-0.083$ between $z\sim2.3$ and $z=0$ are in agreement with our results presented in \S~\ref{comp_local}, that the redshift evolution of the bTFR is non-monotonic.}
    \label{fig:alter-offset}
\end{figure}

We consider an alternative, non-parametric approach. In Figure~\ref{fig:alter-offset} we show our TFR galaxies at $z\sim2.3$ (red) and $z\sim0.9$ (blue) together with the local sample by \cite{Lelli16} (black) in the bTFR plane. In the mass bins labeled `A', `B', and `C', we compute the weighted mean velocity of each redshift and mass subsample. We then compare the weighted mean velocities at different redshifts, as indicated in the figure, and determine an average velocity difference from combining the results from individual mass bins.

Although this approach is strongly limited by the number of galaxies per mass bin, and by the common mass range which is spanned by low- as well as high$-z$ galaxies, its advantage becomes clear: not only is the resulting offset in velocity independent of any functional form usually given by a TFR, but the method would also be sensitive to changes of the TFR slope with redshift if the covered mass range would be large enough.

For our TFR samples, we find an average difference in velocity as measured from the average local velocity minus the average high$-z$ velocity, $\Delta \overline{{\rm log}(v_{\rm circ}~[{\rm km/s}])}$, of $-0.119$ between $z=0$ and $z\sim0.9$, and of $-0.083$ between $z=0$ and $z\sim2.3$. This confirms our result presented in \S~\ref{comp_local}, that the bTFR evolution is not a monotonic function of redshift.

\section{The impact of mass uncertainties on slope and residuals of the TFR}\label{uncertainties}

\begin{figure}
	\centering
	\includegraphics[width=\columnwidth]{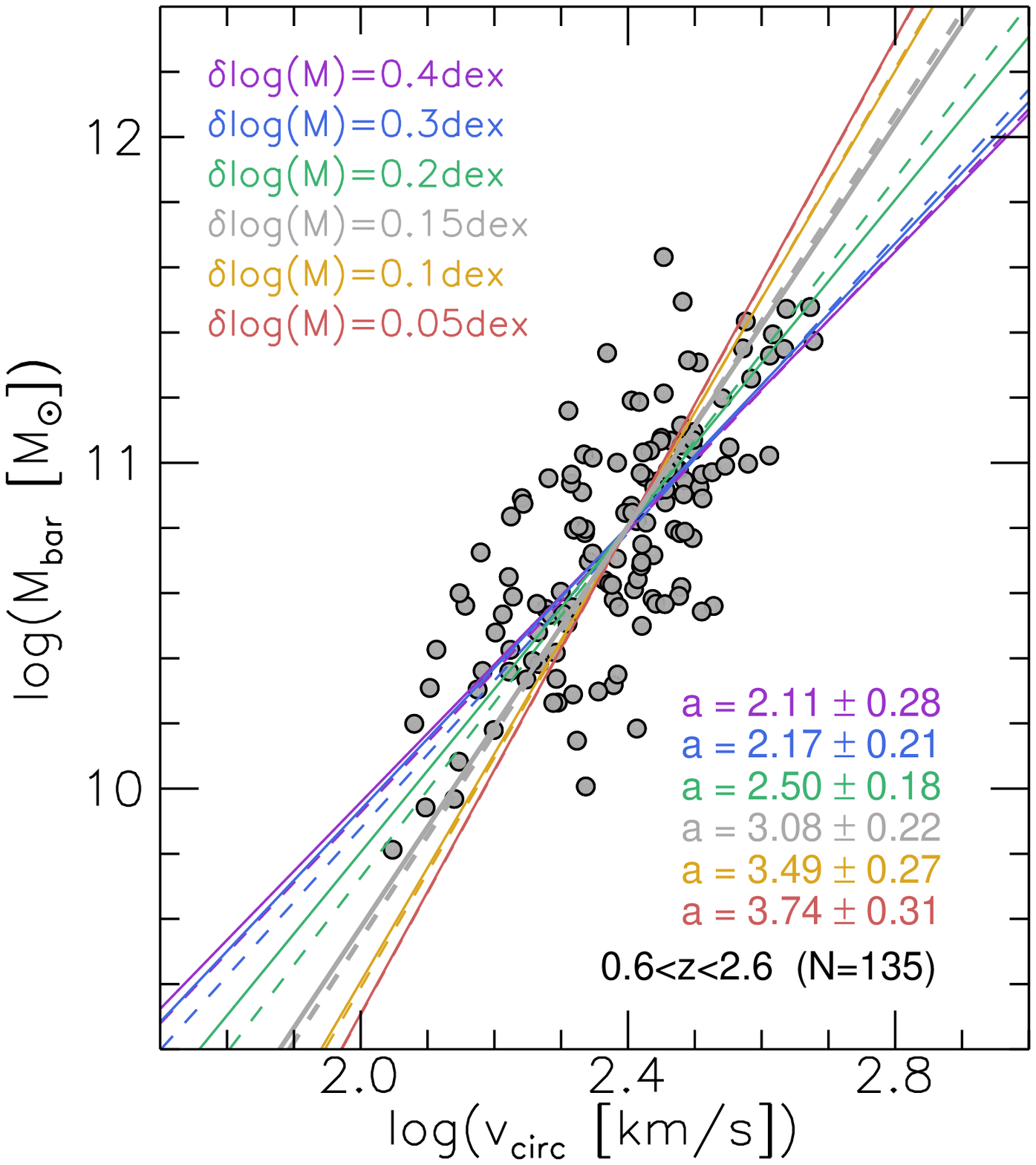}
    \caption{Effect of varying uncertainties for the baryonic mass estimates on the slope of the bTFR for our full TFR sample, as indicated in the legend (solid lines, least-squares fits). 
    The resulting best-fit slopes $a$ vary by a factor of $\sim2$ for the explored range of mass uncertainties. As dashed lines, we show the corresponding fits using the Bayesian approach by \cite{Kelly07} which show a similar behaviour.}
    \label{fig:slopes}
\end{figure}

The slope and scatter of the TFR are affected by the adopted uncertainties in mass. 
In Figure~\ref{fig:slopes} we show fit examples to the bTFR of the full sample with varying assumptions for the mass uncertainties, namely $0.05\leq\delta{\rm log}(M_{\rm bar}~[M_{\odot}])\leq0.4$. The corresponding changes in slope (from $a=2.11$ to $a=3.74$) are well beyond the already large fit uncertainties on the individual slopes, confirming that a proper assessment of the mass uncertainties is essential. 
For simple linear regression, the effect of finding progressively flatter slopes for samples with larger uncertainties is known as `loss of power', or `attenuation to the null' \citep[e.g.][]{Carroll06}.
The relevant quantity for our study, however, is the change in zero-point offset, which is for the explored range only $0.02$~dex. This is due to the use of $v_{\rm ref}$ in Equation~\eqref{eq:fit} which ensures only little dependence of the zero-point~$b$ on the slope~$a$.

Variations of the TFR slope naturally affect the TFR residuals to the best-fit relation \citep[see also][]{Zaritsky14}. We define the TFR residuals as follows:
\begin{equation}\label{eq:residual}
	\Delta{\rm log}(v_{\rm circ})={\rm log}(v_{\rm circ}) - \Bigg[\frac{-b}{a}+\frac{{\rm log}(M/M_{\odot})}{a}+{\rm log}(v_{\rm ref})\Bigg].
\end{equation} 

\begin{figure}
	\includegraphics[width=\columnwidth]{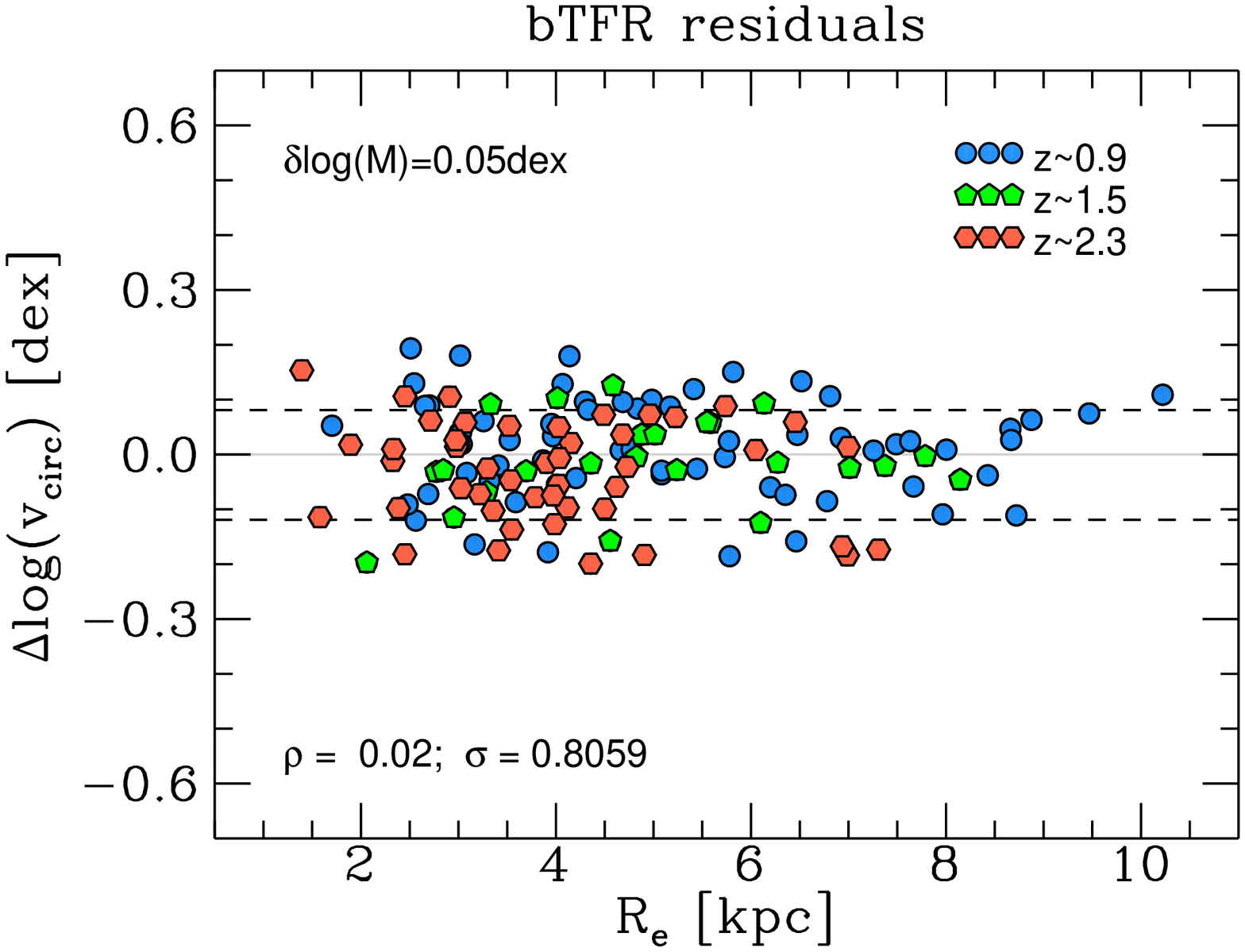}
	\includegraphics[width=\columnwidth]{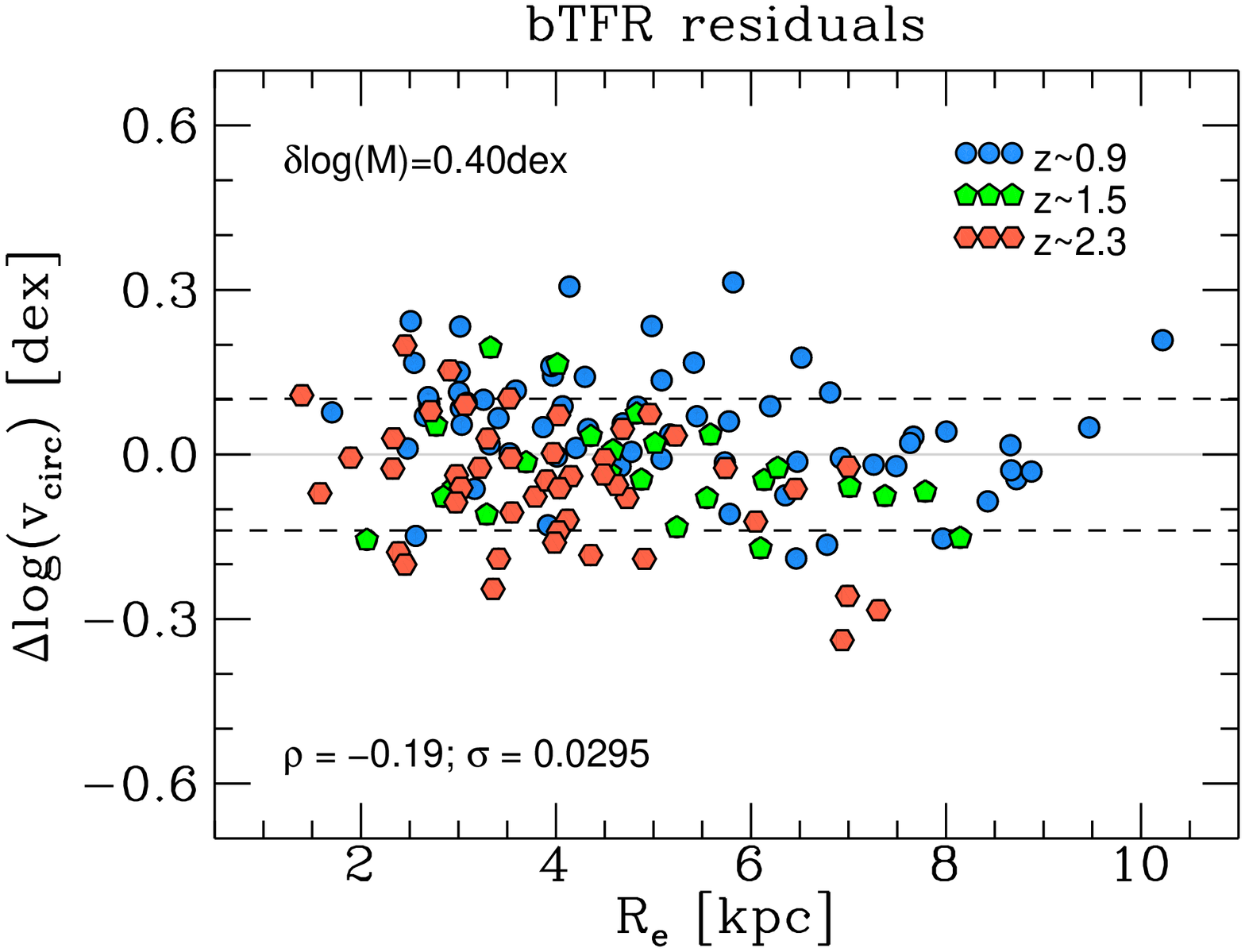}
    \caption{{\bf Top panel:} residuals of the bTFR as a function of effective radius, using $\delta M_{\rm bar}=0.05$~dex. The dashed lines show the sample standard deviation. While we find no significant correlation for our full sample ($\rho=0.02$, $\sigma=0.8059$), a slightly stronger correlation for the highest redshift bin (red) is visible.
    {\bf Bottom panel:} same as above, but using $\delta M_{\rm bar}=0.4$~dex. We find a weak correlation for our full sample ($\rho=-0.19$, $\sigma=0.0295$), and again a slightly stronger correlation for the highest redshift bin.}
    \label{fig:dev_btfr_reff}
\end{figure}

To demonstrate the effect of changing the slope, we show in Figure~\ref{fig:dev_btfr_reff} the bTFR residuals as a function of $R_{e}$. In the upper panel, we show the residuals to a fit with baryonic mass uncertainties of 0.05~dex, leading to a slope which approximately corresponds to the local slope by \cite{Lelli16}. In the lower panel, we show the same for a fit adopting 0.4~dex uncertainties for $M_{\rm bar}$. While there is no correlation found for the former case (Spearman correlation coefficient $\rho=0.02$ with a significance of $\sigma=0.8059$), we find a weak correlation when adopting $\delta M_{\rm bar}=0.4$~dex ($\rho=-0.19$, $\sigma=0.0295$).

\begin{figure}
	\includegraphics[width=\columnwidth]{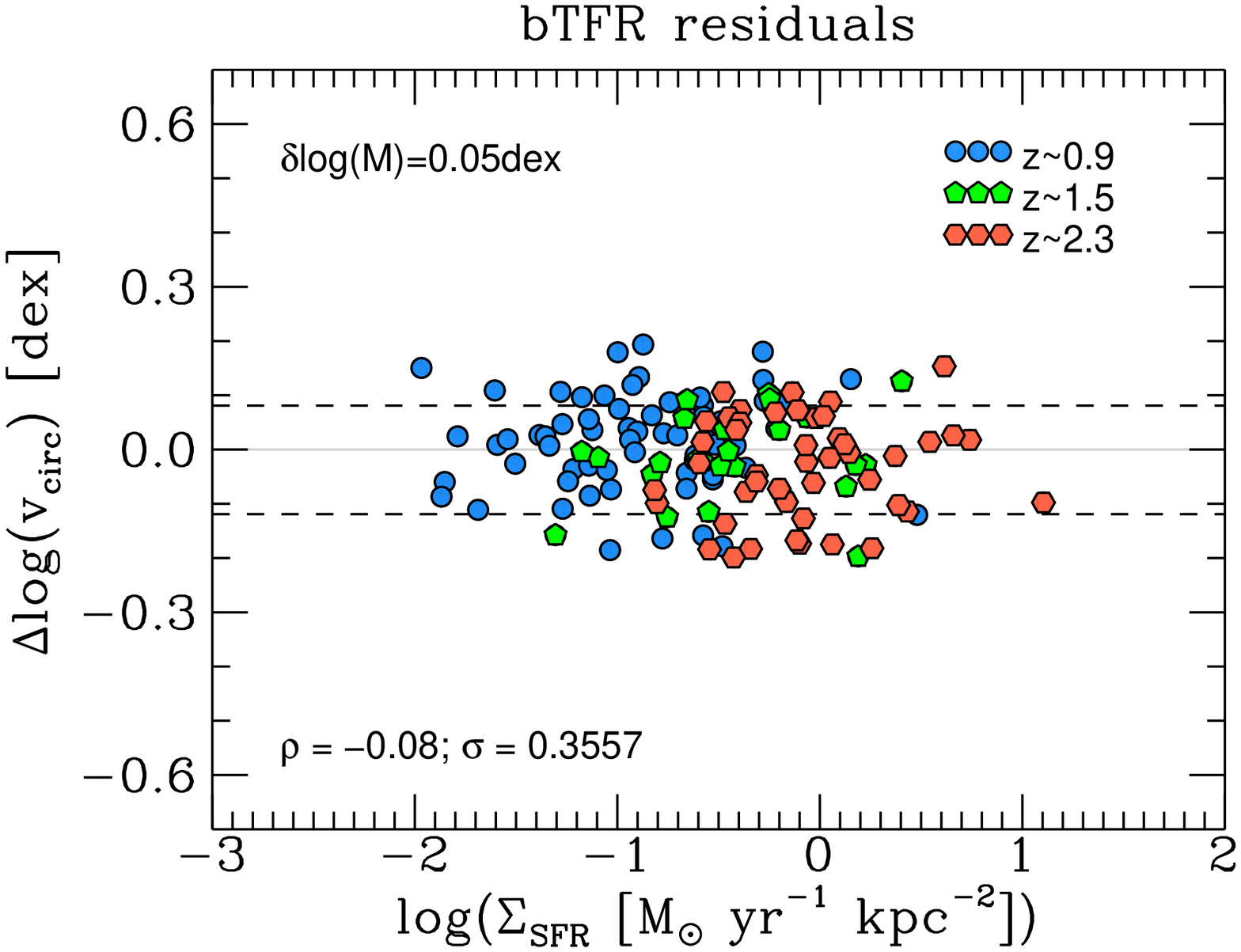}
	\includegraphics[width=\columnwidth]{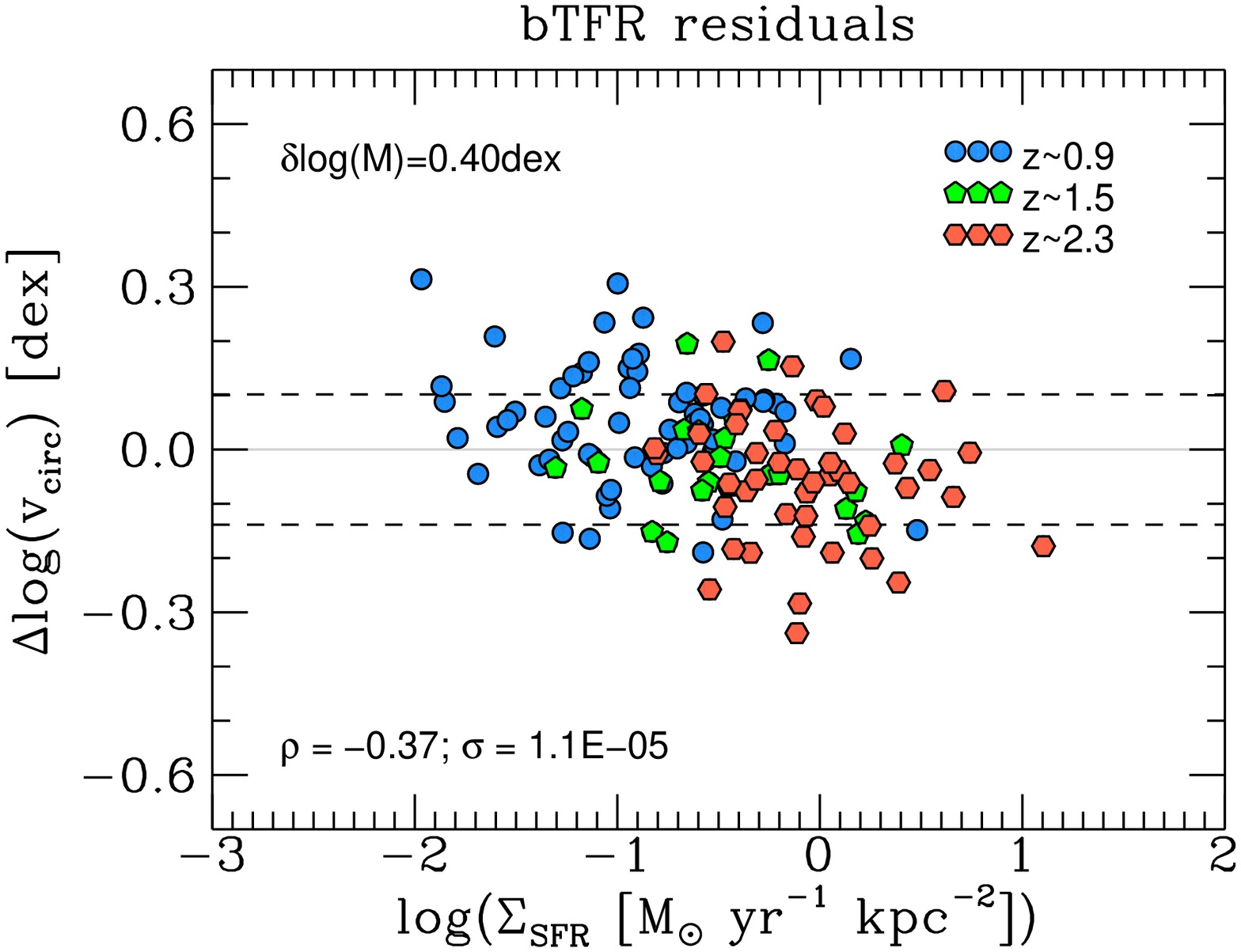}
	\caption{{\bf Top panel:} residuals of the bTFR as a function of SFR surface density $\Sigma_{\rm SFR}$, using $\delta M_{\rm bar}=0.05$~dex. The dashed lines show the sample standard deviation. We find no correlation for our fiducial fit ($\rho=-0.08$, $\sigma=0.3557$).
	{\bf Bottom panel:} same as above, but using $\delta M_{\rm bar}=0.4$~dex. We find a significant correlation ($\rho=-0.37$, $\sigma=1.1\times10^{-5}$).}
    \label{fig:dev_btfr_SSFR}
\end{figure}

We find a similar behaviour for baryonic (and stellar) mass surface density, with no significant correlation between TFR offset and mass surface density for the $\delta M_{\rm bar}=0.05$~dex fit, but a strong correlation for the $\delta M_{\rm bar}=0.4$~dex fit (not shown). 
No correlation for the $\delta M_{\rm bar}=0.05$~dex fit residuals is found for SFR surface density ($\rho=-0.08$, $\sigma=0.3557$), but a significant correlation with $\rho=-0.37$ and $\sigma=1.1\times 10^{-5}$ for the $\delta M_{\rm bar}=0.4$~dex fit (Figure~\ref{fig:dev_btfr_SSFR}). 

From this exercise it becomes clear that the high$-z$ slope, and with it the TFR residuals, are strongly dependent on the accuracy of the mass and SFR measurements.

\section{Derivation of the toy model for TFR evolution}\label{model-derivation}

\begin{figure*}
\centering
	\includegraphics[width=0.8\textwidth]{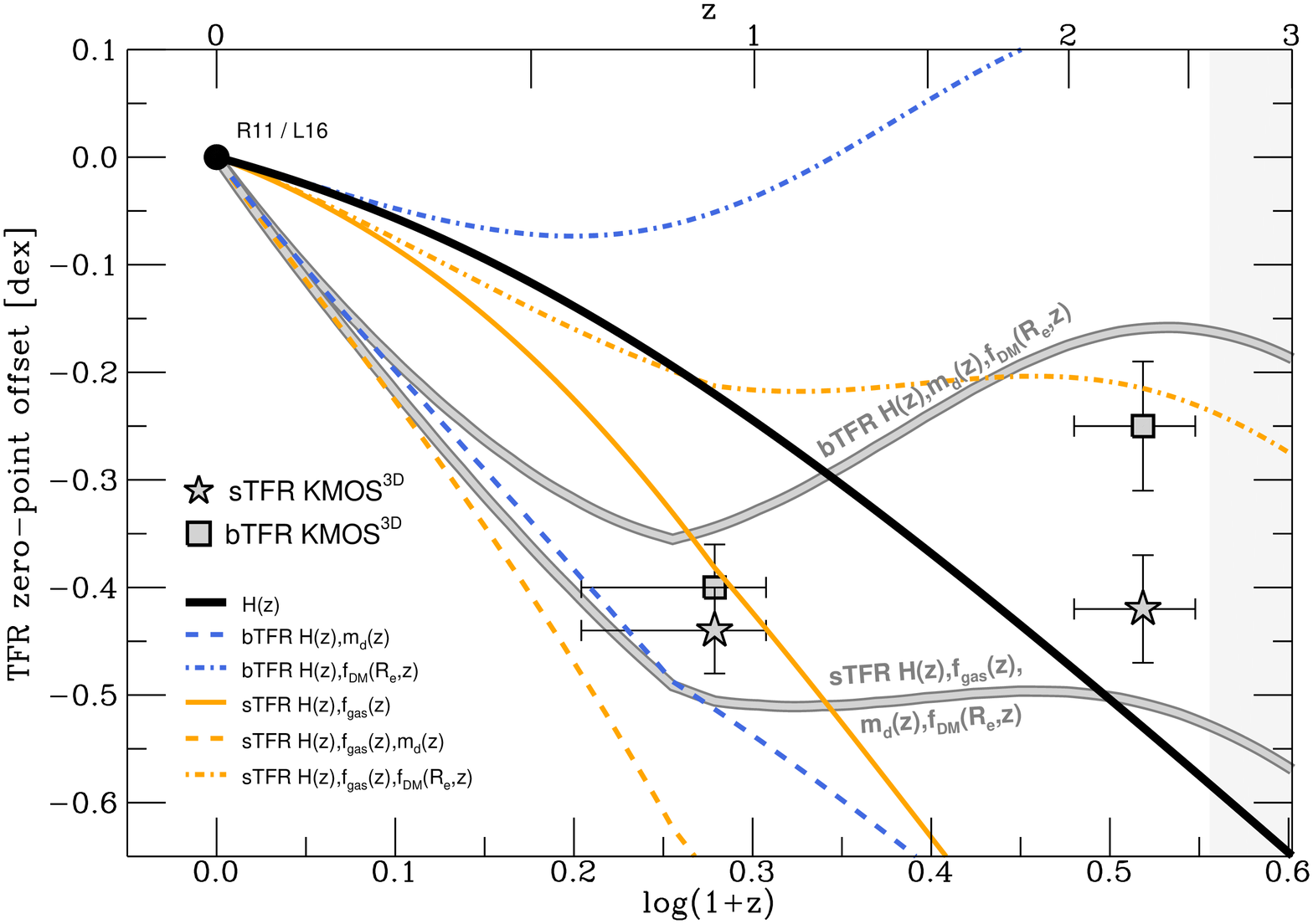}
    \caption{TFR zero-point offsets of the stellar and baryonic mass TFRs as a function of cosmic time. The symbols show the \kd data in relation to the corresponding local normalizations by \citeauthor{Reyes11}\ (2011; R11) and \citeauthor{Lelli16}\ (2016b; L16), as shown in Figure~\ref{fig:context}. 
    The black line shows the TFR evolution for a model governed solely by $H(z)$.
    The colored lines show toy models for the bTFR (blue) and the sTFR (orange) evolution for different combinations of additional redshift dependencies of $f_{\rm gas}$, $f_{\rm DM}(R_{e})$, or $m_{d}$, as detailed in Appendix~\ref{model-derivation}, and as indicated in the legend. The grey lines show our final toy model following Equations~\eqref{eq:btfr} and \eqref{eq:stfr} and including $f_{\rm gas}(z)$, $f_{\rm DM}(R_{e},z)$, and $m_{d}(z)$ as shown in inset (a) in Figure~\ref{fig:context}.}
    \label{fig:tfrev_effects}
\end{figure*}

\subsection{The theoretical framework}\label{model-theory}

In the following, we give details on the theoretical toy model derivation of the TFR and its evolution. The relationship between the DM halo mass, radius, and circular velocity are given by Equations~\eqref{eq:virial}, describing a truncated isothermal sphere.
A plausible model for a SFG which has formed inside the dark halo is a self-gravitating thin baryonic disk with an exponential surface density profile 
\begin{equation}\label{eq:surface}
	\Sigma(r)=\Sigma_{0}\,e^{-r/R_{d}},
\end{equation}
where $\Sigma_{0}$ is the central surface density, related to the baryonic disk mass as $M_{\rm bar}\propto\Sigma_{0}\,R_{d}^{2}$. In reality, disk galaxies feature a finite thickness. This does not affect the scalings presented here \citep[see e.g.][and references therein]{Courteau99, BT08}.
To associate the baryonic disk to the dark halo, one can assume a simple model where the corresponding masses and radii are related through a proportionality factor:
\begin{equation}\label{eq:md}
	M_{\rm bar}=m_{d}\cdot M_{h} \hspace{3mm} {\rm ;} \hspace{3mm} R_{\rm bar}=r_{f} \cdot R_{h}.
\end{equation}
$R_{\rm bar}$ can be expressed through the disk scale length $R_{d}$, or the effective radius $R_{e}$, which for rotation-dominated disks are related through $R_{e}\approx R_{d}\cdot 1.68$. As noted in Section~\ref{conclusions}, we take $r_{f}$ to be independent of redshift. In standard models of disk galaxy evolution, $r_{f}$ combines information on the halo spin parameter, on the halo concentration parameter, and on the ratios of the angular momenta and masses of baryons and DM (cf.\ Equation~(28) of \cite{MMW98}, accounting for adiabatic contraction). It has however been shown that the ratio between $R_{h}$ and $R_{d}$ is approximately constant for massive SFGs in the redshift range $0.8<z<2.6$ \citep{Burkert16}. This does also hold for our TFR sample and the average values at $z\sim0.9$ and $z\sim2.3$, even though there is substantial scatter for individual objects. 

To quantify the contributions of baryons and DM to the circular velocity at a given radius we write 
\begin{equation}\label{eq:vc}
	v_{\rm circ}(r)=\sqrt{v_{\rm bar}^{2}(r)+v_{\rm DM}^{2}(r)}.
\end{equation}
The baryonic contribution can be computed, for instance, using the expression for an infinitely thin exponential disk \citep{Freeman70},
\begin{equation}\label{eq:freeman}
	v_{\rm bar}^{2}(r)=4\pi G\,\Sigma_{0}R_{d}y^2[I_{0}(y)K_{0}(y)-I_{1}(y)K_{1}(y)],
\end{equation}
where $y=r/(2R_{d})$, and $I_{i}(y)$ and $K_{i}(y)$ are the modified Bessel functions of the first and second kind.
At $r=R_{e}$, this equation becomes
\begin{equation}\label{eq:freeman_Re}
	v_{\rm bar}^{2}(R_{e})=\frac{M_{\rm bar}}{R_{d}}\cdot C^{\prime\prime},
\end{equation}
where $C^{\prime\prime}$ is a constant.
The DM component can be derived simply through a DM fraction at the radius of interest, $f_{\rm DM}(r)=v_{\rm DM}^2(r)/v_{\rm circ}^2(r)$, or via adopting a full mass profile \citep[e.g. NFW or Einasto,][]{NFW96, Einasto65}.

Equations~\eqref{eq:virial} can be combined to 
\begin{equation}\label{eq:comb}
	M_{h}= R_{h}^3 H(z)^2~10^{2}~G^{-1}.
\end{equation}
By inserting Equations~\eqref{eq:md} into Equation~\eqref{eq:comb}, and by substituting $R_{d}$ through a re-arranged Equation~\eqref{eq:freeman_Re}, one arrives at Equation~\eqref{eq:btfr} given in Section~\ref{conclusions}. After introducing the gas fraction $f_{\rm gas}=M_{\rm gas}/M_{\rm bar}$, one arrives at Equation~\eqref{eq:stfr}.
These equations predict a TFR evolution with a constant slope, but evolving zero-point with cosmic time, depending not only on $H(z)$, but also on changes in $m_{d}$, $f_{\rm DM}(R_{e})$, and $f_{\rm gas}$ with cosmic time.

We note that deviations from the proposed slope ($a=3$) can be related to additional dependencies on $v_{\rm bar}$, e.g.\ of the surface density $\Sigma$ \citep{Courteau07}.

\subsection{Observational constraints on the redshift evolution of $f_{\rm gas}$,  $m_{d}$, and $f_{\rm DM}(R_{e})$}\label{model-observations}

In the following paragraphs, we discuss the motivation for the adopted redshift evolution of $f_{\rm gas}$,  $m_{d}$, and $f_{\rm DM}(R_{e})$ in the toy model context.
Figure~\ref{fig:tfrev_effects} summarizes the individual and combined effects of adopting the respective redshift evolutions of $f_{\rm gas}$, $m_{d}$, and $f_{\rm DM}(R_{e})$ for the bTFR and sTFR evolution.

\subsubsection{The redshift evolution of $f_{\rm gas}$}\label{gas_ev}

For our toy model approach, we consider the gas fraction $f_{\rm gas}$ to be the sum of molecular and atomic gas mass divided by the total baryonic mass, $f_{\rm gas}=(M_{\rm gas, mol}+M_{\rm gas, at})/(M_{\rm gas, mol}+M_{\rm gas, at}+M_{*})$.
The evolution of the molecular gas mass-to-stellar mass ratio is given through the scaling relation by \cite{Tacconi17}:
\begin{equation}
\begin{aligned}
	{\rm log}\left(\frac{M_{\rm gas,mol}}{M_{*}}\right)\approx &\, 0.12 - 3.62\cdot\left[{\rm log}(1+z)-0.66\right]^2 \\ 
	& - 0.33\cdot\left[{\rm log}(M_{*}~[M_{\odot}])-10.7\right].
\end{aligned}
\end{equation}
Here, we do not take into account the additional dependencies given in the full parametrization by \citet{Tacconi17} on MS offset, and offset from the M-R relation, but assume that the model galaxies lie on these relations.

Locally, the galactic gas mass is dominated by atomic gas. To account for atomic gas mass at $z=0$, we use the fitting functions presented by \cite{Saintonge11}. We use a local reference stellar mass of log$(M_{*}~[M_{\odot}])=10.94$, i.e.\ the stellar mass corresponding to our reference velocity $v_{\rm ref}=242$~km/s in the context of the sTFR fit by \cite{Reyes11}.

To account for atomic gas masses at $z>0$, we follow the theoretical prediction that, at fixed $M_{*}$, the ratio of atomic gas mass to stellar mass does not change significantly with redshift \citep[e.g.][]{Fu12}. We use again the fitting functions by \cite{Saintonge11} to now determine the atomic gas mass for galaxies with log$(M_{*}~[M_{\odot}])=10.50$, which corresponds to the average stellar mass of our TFR galaxies at $v_{\rm ref}=242$~km/s in both redshift bins.

Between $z=0$ and $z=0.9$, we assume a smooth TFR evolution, meaning that at fixed circular velocity, galaxies have decreasing $M_{*}$ with increasing redshift, in order to compute the gas fractions. Although we cannot quantify this assumption with our observations, we note that in comparing to our data, only the relative offset in $f_{\rm gas}$ (or any other parameter discussed below) between $z=0$, $z=0.9$, and $z=2.3$ is relevant. Our assumption therefore serves mainly to avoid sudden (unphysical) offsets in the redshift evolution of $f_{\rm gas}$.

Corresponding values of the gas mass fraction at $z=\{0.0; 0.9; 2.3\}$ are $f_{\rm gas}\approx\{0.07; 0.36; 0.58\}$.

\subsubsection{The redshift evolution of $m_{d}$}\label{md_ev}

The baryonic disk mass fraction, $m_{d}=M_{\rm bar}/M_{h}$, is not a direct observable, since it depends on the usually unknown DM halo mass.
For the local Universe, we use the fitting function by \cite{Moster13} from abundance matching to determine a stellar disk mass fraction, $m_{d,*}=M_{*}/M_{h}$. For a stellar mass of log$(M_{*}~[M_{\odot}])=10.94$, this gives $m_{d,*}\approx0.012$. Again, we use the fitting functions by \cite{Saintonge11} to determine the corresponding gas mass, taking into account contributions from helium via $M_{\rm He}\approx0.33~M_{\rm H{\textsc i}}$. This results in a baryonic disk mass fraction at $z=0$ of $m_{d}\approx0.013$.

The recent study by \cite{Burkert16} finds a typical value of $m_{d}=0.05$ for SFGs at $0.8<z<2.6$ based on a Monte-Carlo NFW modelling of data from the \kd and SINS/zC-SINF \citep{FS09, Mancini11} surveys. These galaxies have masses similar to the galaxies in our TFR sample. We adopt their value of $m_{d}=0.05$ for $0.8<z<2.6$.

Between $z=0$ and $z=0.8$ we assume a linear increase of $m_{d}$. Clearly, this is a simplifying conjecture. As for the atomic gas masses, we emphasize that this assumption has primarily cosmetic effects, while the crucial quantity is the relative difference in $m_{d}$ between $z=0$, $z\sim0.9$, and $z\sim2.3$.

\subsubsection{The redshift evolution of $f_{\rm DM}(R_{e})$}\label{dm_ev}

For the DM fraction of local disk galaxies, we follow Figure~1 by \cite{Courteau15} which, among others, shows galaxies from the DiskMass survey \citep{Martinsson13a, Martinsson13b}. At $v_{\rm circ}=242$~km/s, DM fractions of local disk galaxies lie roughly between $f_{\rm DM}(r_{2.2})=0.55$ and $f_{\rm DM}(r_{2.2})=0.75$, with large scatter and uncertainties.

At higher redshift, \citetalias{WuytsS16} derived DM fractions from the difference between dynamical and baryonic masses of the \kd subsample of 240 SFGs, which represents our parent sample. Corresponding values, also corrected for mass completeness, are given in their Table~1.

For convenience, we parametrize the evolution of the DM fraction with redshift as follows: $f_{\rm DM}(R_{e})=0.7 \cdot {\rm exp}[-(0.5\cdot z)^{2.5}]$.
This gives an evolution which is somewhat stronger than what is suggested by just taking the average values provided by \cite{Courteau15} and \citetalias{WuytsS16}, but easily within the uncertainties presented in both papers. 
We adopt this marginally stronger evolution to better match our observed TFR offsets with the toy model.

Corresponding values of the DM fraction at $z=\{0.0; 0.9; 2.3\}$ are $f_{\rm DM}(R_{e})\approx\{0.70; 0.61; 0.17\}$.

We note that our toy model evolution is particularly sensitive to the parametrization of $f_{\rm DM}(R_{e},z)$ which is in our implementation with the simplistic description for $m_{d}(z)$ responsible for the flattening/upturn of the sTFR/bTFR (see Figure~\ref{fig:tfrev_effects}). The high value for the local DM fraction (which would at $r=R_{e}$ rather be lower than at $r=r_{2.2}$) as well as the comparably strong evolution at $z>1$ can certainly be challenged.

\subsubsection{Comments on the evolution of the halo concentration parameter}\label{c_ev}

The predicted evolution of the halo concentration parameter $c$ between $z=2$ and $z=0$ for haloes of masses that are relevant to this study (i.e.\ central stellar masses of log$(M_{*}~[M_{\odot}])\approx 10.5$ at $z\sim2$, and log$(M_{*}~[M_{\odot}])\approx 10.9$ at $z\sim0$) goes from $c\approx4$ at $z=2$ to $c\approx6$ at $z=1$ and to $c\approx7$ at $z=0$ \citep{Dutton14}. This alone would increase the DM fraction at $R_{e}$ by roughly $0.1$. 

Starting from the central DM fractions as determined by \citetalias{WuytsS16}, abundance-matched haloes \citep{Moster13} would require concentrations of $c\approx3$ and $c\approx12$ at $z\sim2.3$ and $z\sim0.9$, respectively \citep[cf.\ Eq.~19 by][]{MMW98}. Extending this to $z=0$ is not straight-forward since local late-type galaxies have typically lower circular velocity as required for the extrapolation of the local TFR to our $v_{\rm ref}=242$~km/s (see discussion in \S~\ref{comp_local}). However, using the stellar mass-radius relation presented by \cite{vdWel14a}, inferred concentrations of these hypothetical haloes would have to be $c\approx13$.

This points towards a potential issue in the observational constraints to our toy model because the $m_{d}$ values inferred by \cite{Burkert16} are based on Monte-Carlo modelling involving standard NFW haloes. One could consider fitting $m_{d}$ to better match the observed TFR zero-point evolution.

In general, the possible effects of adiabatic contraction or expansion of the halo as a response to baryonic disk formation make theoretical predictions of the central DM fractions uncertain (see e.g.\ the discussions by \citealp{Duffy10, Velliscig14, Dutton16}; and also \citealp{Dutton14} for an overview of predictions of concentration-mass relations from analytical models).

\section{Physical properties of galaxies in the TFR sample}\label{ind_prop}

In Table~\ref{tab:ind-properties} we list redshift $z$, stellar mass $M_{*}$, baryonic mass $M_{\rm bar}$, maximum modelled circular velocity $v_{\rm circ, max}$, and modelled intrinsic velocity dispersion $\sigma_{0}$ of our TFR galaxies. The full table is available in machine readable form.

\begin{longtable*}{rrrrrr}
 \caption{Physical properties of galaxies in our TFR sample in terms of redshift $z$, stellar mass $M_{*}$, baryonic mass $M_{\rm bar}$, maximum modelled circular velocity $v_{\rm circ, max}$, and modelled intrinsic velocity dispersion $\sigma_{0}$.}\label{tab:ind-properties}\\\hline
   \# & $z$ & log($M_{*}~[M_{\odot}]$) & log($M_{\rm bar}~[M_{\odot}]$) & $v_{\rm circ, max}$~[km/s] & $\sigma_{0}$~[km/s] \\\hline
\endfirsthead
\caption* {\textbf{(Continued)}}\\\hline
  \# & $z$ & log($M_{*}~[M_{\odot}]$) & log($M_{\rm bar}~[M_{\odot}]$) & $v_{\rm circ, max}$~[km/s] & $\sigma_{0}$~[km/s] \\\hline
\endhead
\endfoot
\endlastfoot
  1 & 0.602 & 10.85 & 10.93 & 274.9 &  30.9 \\
  2 & 0.626 & 11.00 & 11.07 & 314.3 &  25.8 \\
  3 & 0.669 & 10.76 & 10.82 & 267.5 &  49.8 \\
  4 & 0.678 & 10.49 & 10.58 & 273.4 &  38.5 \\
  5 & 0.758 & 10.66 & 10.77 & 313.8 &  24.3 \\
  \vdots & \vdots & \vdots & \vdots & \vdots & \vdots \\
 \hline
\end{longtable*}

\bibliographystyle{aasjournal.bst}
\bibliography{TFRliterature}



\end{document}